\documentclass{aa}

\usepackage{booktabs, makecell, multirow} 

\usepackage{graphicx}
\usepackage{txfonts}
\usepackage{color}
\usepackage{verbatim}
\usepackage{subcaption}         
\usepackage{lscape}             
\usepackage{placeins}           
                                
\begin{document}



   


   
    

\title{Scattering and sputtering on the lunar surface}
\subtitle{Insights from negative ions observed at the surface}


\author{
   Romain Canu-Blot\inst{\ref{inst1},\ref{inst2}}
    \and Martin Wieser\inst{\ref{inst1}}
    \and Umberto Rollero\inst{\ref{inst1},\ref{inst2}}
    \and Thomas Maynadié\inst{\ref{inst1},\ref{inst2}}
    \and Stas Barabash\inst{\ref{inst1}}
    \and Gabriella Stenberg Wieser\inst{\ref{inst1}}
    \and Aibing Zhang \inst{\ref{inst3},\ref{inst4}}
    \and Wenjing Wang \inst{\ref{inst3}}
    \and Chi Wang \inst{\ref{inst3},\ref{inst4}}
         }     

   \institute{Swedish Institute of Space Physics (IRF), Bengt Hultqvists väg 1, Kiruna, 98192, Sweden\\
             \email{romain.canu-blot@irf.se}\label{inst1}
             \and
             Department of Physics, Ume{\aa} University, Ume{\aa}, Sweden\label{inst2}
             \and
             National Space Science Center (NSSC), Chinese Academy of Sciences (CAS), No. 1 Nanertiao, Zhongguancun, Haidian district, Beijing, 100190, China\label{inst3}
             \and
             University of Chinese Academy of Sciences, No. 1 Yanqihu East Road, Huairou district, Beijing, 101408, China\label{inst4}
}



   \date{Received February 18, 2026}

 \abstract
    {Airless planetary bodies are directly exposed to solar wind ions, which can scatter or become implanted upon impact with the regolith-covered surface, while also sputtering surface atoms.}
   {We construct a semi-analytical model for the scattering of ions of hundreds of eV and the sputtering of surface atoms, both resulting in the emission of negative ions from the lunar surface. Our model contains a novel description of the scattering process that is physics-based and constrained by observations.}
   {We use data from the Negative Ions at the Lunar Surface (NILS) instrument on the Chang’e-6 lander to update prior knowledge of ion scattering and sputtering from lunar regolith through Bayesian inference.}
   {Our model shows good agreement with the NILS data. A precipitating solar wind proton has roughly a 22\% chance of scattering from the lunar surface in any charge state, and about an 8\% chance of sputtering a surface hydrogen atom. The resulting ratio of scattered to sputtered hydrogen flux is $\eta^{sc}/\eta^{sp} = 1.5^{+1.5}_{-1.1}$ for a proton speed of 300\,km/s. We find a high probability (7–20\%) that a hydrogen atom leaves the surface negatively charged. The angular emission distributions at near-grazing angles for both scattered and sputtered fluxes are controlled by surface roughness. Our model also indicates significant inelastic energy losses for hydrogen interacting with the regolith, suggesting a longer effective path length than previously assumed. Finally, we estimate a surface binding energy of 5.5\,eV, consistent with the observations.
    }
   {Our model describes the scattering and sputtering of particles of any charge state from any homogeneous, multi-species surface. Using NILS data, we successfully applied the model to update our understanding of solar wind interacting with lunar regolith, and the emission of negative hydrogen ions.}


\maketitle

\nolinenumbers


\section{Introduction}

Planetary bodies without atmospheres are directly exposed to the space environment. Continuous space weathering \citep{Pieters_2016} alters their surfaces and produces a porous, unconsolidated and fragmented layer of debris known as regolith \citep{Papike_1982, Clark_2002}. The regolith acts as the primary interface between airless bodies and the surrounding space environment, controlling how solar wind and other energetic particles interact with their surfaces. Understanding these interactions is essential for studying surface evolution, elemental composition, and the formation of surface-bounded exospheres.

When a particle interacts with a solid surface, it may be directly reflected by surface atoms or penetrate into the solid. Within the solid, the particle can scatter off atoms, lose energy, and eventually exit the material. Alternatively, it may come to rest and be absorbed. The reflection of incoming particles is referred to as "(back)scattering." As the particle moves through the material, it transfers energy to atoms of the solid. Some of these energized atoms may be ejected from the surface--a process known as "sputtering."

The charge state of particles interacting with a surface is not conserved. Precipitating ions are neutralized when approaching the surface and their original charge state is effectively lost \citep[for example,][]{jans2001_bazro3}. Near the surface, particles remain in a (quasi-)neutral charge-state equilibrium that is established on time scales much shorter than the total interaction time with the material \citep{Lienemann_2011}. As particles leave the surface, their charge states are further modified through charge-exchange processes, eventually reaching a final charge state at distances of several nanometres from the solid. The probability for an emitted particle to reach a particular charge state depends on the energy and species of the particle, but also on surface properties like average composition, adsorbents, atomic structure (crystalline or amorphous), structure defects, or composition impurities \citep[for example,][]{Kawano:1983aa, Maazouz_1998, Wucher2008, Wucher2013,Cartry:2017aa}.

The processes of scattering, sputtering, and charge-exchange have been extensively studied for a wide range of particle-surface combinations. However, because the regolith covering airless bodies is unique, studies of natural, or pristine, regolith--as opposed to returned samples, simulants or simulations--are limited to in-situ or space-borne observations.

Among airless bodies, the Moon provides the most accessible example of natural regolith exposed directly to the solar wind, and much of our empirical understanding is derived from orbital measurements. The lunar surface is entirely covered by regolith composed primarily of silicate minerals and glassy agglutinates formed by impact melting \citep{Heiken1975Petrology-of-lu}. Most solar wind protons impacting the lunar regolith are absorbed, sourcing the surface with hydrogen and hydroxyl-bearing molecules \citep{Tucker_2019}. Approximately 10–20\% of incident protons result in the emission of neutral hydrogen with energies greater than 10\,eV \citep{McComas_2009, Wieser_2009, Schaufelberger_2011,  RodriguezM._2012,  Allegrini_2013,  Saul_2013, Vorburger_2014, Zhang_2020}. These neutrals are produced either by scattering of the incoming protons or by sputtering of hydrogen from the surface, in both cases followed by neutralization of the emitted particle. In addition to neutrals, about 0.1–1\% of solar wind protons backscatter from the lunar surface as protons \citep{Saito_2008, Lue_2014, Lue_2018} and about 2.5\% as negative hydrogen ions \citep{Wieser2025}. Beyond hydrogen, a wide selection of likely sputtered heavier positive ions was observed \citep{Tanaka_2009, Yokota_2009}, along with other energetic neutral atoms including backscattered helium and sputtered oxygen \citep{Vorburger2014oxygenhelium}.

Angular emission functions and the corresponding energy spectra of emitted particles have been reconstructed for energetic neutral hydrogen \citep{Schaufelberger_2011, lue2016scattering} and for protons \citep{Lue:2018aa}, both using data from orbiting spacecraft. Although this enables imaging of magnetic anomalies \citep{Wieser2010_ena_image,Vorburger2012_Energetic-neutr}, the large observation distance from the surface complicates the study of the actual surface interaction. Indeed, orbital observations average over large surface regions of the Moon, blurring small scale local variations of the precipitating solar wind flux caused by magnetic anomalies \citep{Halekas2008_Solar-wind-inte}. This in turn hides the detailed structure of the angular and energy distributions of emitted particles. 

Observations in space are complemented by laboratory experiments and computer simulations. Because access to pristine lunar regolith is limited, laboratory studies are based on returned samples and regolith simulants \citep[for example,][]{Meyer2011Sputtering-of-lr}. In parallel, analytical models and numerical simulations allow one to study particle transport and emission processes under controlled conditions \citep[for example,][]{Toussaint:2017aa, Szabo_2022a, Tucker_2019, Szabo_2023}.

Direct measurements at the lunar surface avoid the material limitations of laboratory studies, the difficulties of accurately modelling regolith in numerical simulations, and the large spatial averaging inherent in observations from orbiting spacecraft. However, only a few instruments have been deployed to date. The Advanced Small Analyzer for Neutrals (ASAN) instrument \citep{Wieser_2020ASAN} measured neutral particles emitted from the lunar surface, and a semi-analytical model was developed to describe the energy distribution of scattered and sputtered neutral hydrogen atoms \citep{Wieser_2024}. In June 2024, the Negative Ions at the Lunar Surface (NILS) instrument \citep{CanuBlot2025}, flown onboard of the Chinese Chang'e-6 mission \citep{Zeng:2023aa}, provided data for negative ions and electrons emitted from the lunar surface. 

In this study, we use NILS data to develop an analytical model for hydrogen scattering and hydrogen sputtering from lunar regolith, including the probability of emission as a negative ion. The model separates the energy and angular distributions and is constrained by direct measurements from the lunar surface.

\section{Instrumentation}

The Negative Ions at the Lunar Surface (NILS) is a small negative ion and electron mass-analyzer \citep{CanuBlot2025} that was flown on the Chinese Chang'e-6 mission to the far-side of the moon \citep{Zeng:2023aa}.

NILS measured direction-resolved energy and mass spectra of electrons and ions with energies ranging from 3\,eV/q to 3\,keV/q. NILS is a single-pixel instrument and can measure only one energy setting and one viewing direction at a time. A complete measurement cycle--scanning all energy and direction settings and transmitting the data to the spacecraft--takes about 100\,s.

The energy range is divided into 48 logarithmically-distributed energy bins. The field of view is an angular slice of $120^\circ \times 20^\circ$, linearly divided into 16 individual viewing directions with different elevation angles relative to the local horizon \citep[Fig.~17]{CanuBlot2025}. An electron-suppression magnet separates ions from the much more abundant electrons. In ion mode, the instrument has a moderate mass resolution of $m/\Delta m \approx 2$, which allows the separation of negative hydrogen and oxygen ions. The particle mass is determined from a velocity measurement in a time-of-flight cell combined with the known energy-per-charge selected by the electrostatic analyser.

The NILS science data consist of a time series of three-dimensional energy–direction–mass matrices \citep{Wieser2025}. Each matrix contains the number of particles detected during a single measurement cycle. The matrices consist of 48 energy bins, 16 elevation bins, and 64 mass bins, which together define the energy-per-charge, elevation angle, and time-of-flight of the measured particle.

\section{Data}

Chang’e-6 landed on 1 June 2024 at 22:23 UTC on the lunar far-side at 153.98°W, 41.64°S. The landing site is close to the large magnetic anomaly cluster in the South Pole-Aitken basin. NILS was operated during the surface-segment of the Chang'e-6 mission. It recorded particle data from the lunar surface for a total of 302 minutes between 2 June 2024 03:01 UTC and 3 June 2024 03:38 UTC, and successfully detected negative ions on the lunar surface \citep{Wieser2025}. 

\begin{figure}[h!]
 \resizebox{0.9\hsize}{!}{\includegraphics{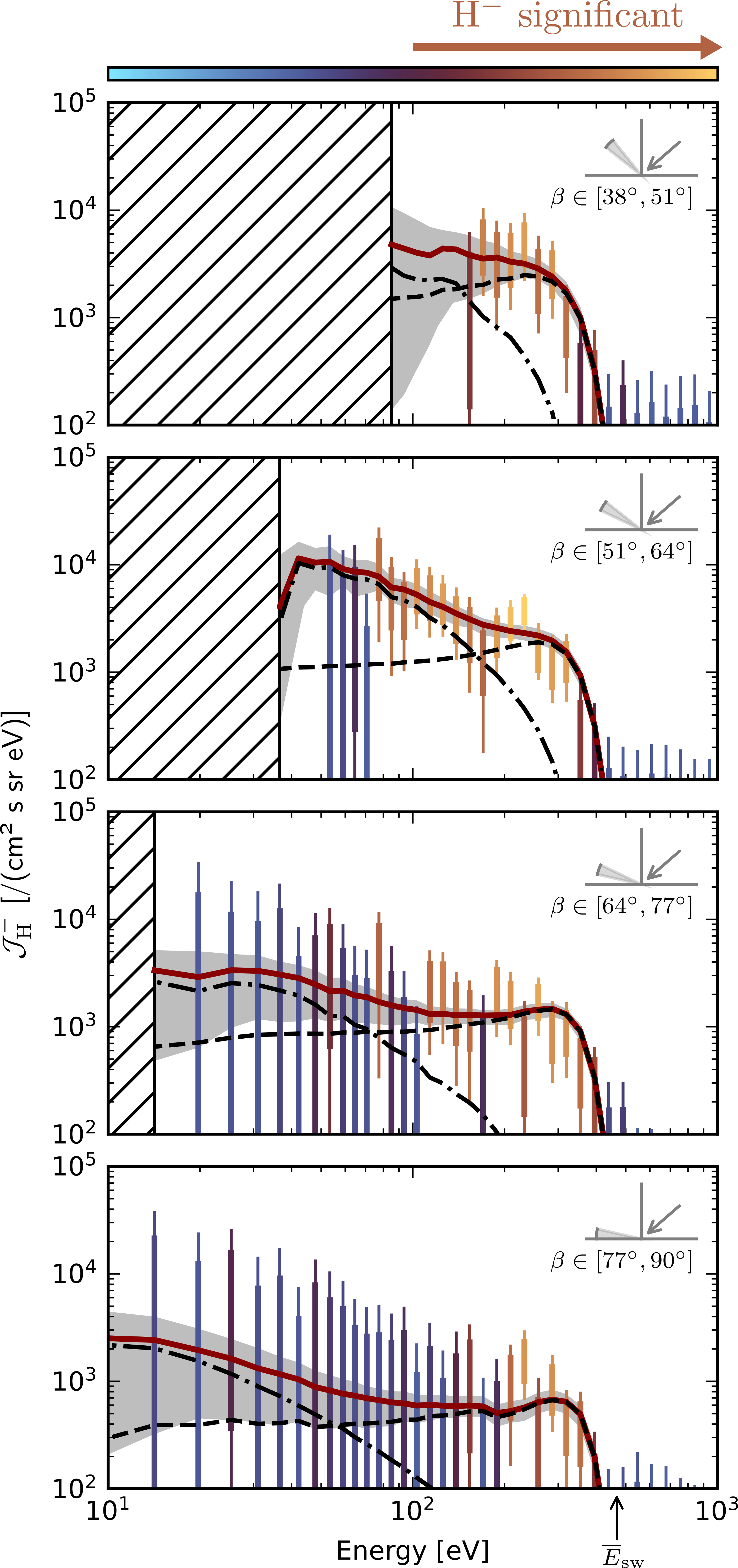}}
  \caption{Differential number flux of negative hydrogen ions, $\mathcal{J}^-_\mathrm{H}$, versus emission energy. Vertical bars show flux estimates, with thick and thin bars representing the 68\% and 90\% highest density intervals, respectively. Bar colour qualitatively reflects the signal significance, based on the Widely Applicable Information Criterion \citep{Watanabe2010} comparing models with and without hydrogen. Each panel corresponds to a specific emission polar angle interval $\beta$ (shown in the upper-right inset), where inward arrows indicate the average solar zenith angle. The energy-axis arrow marks the average undisturbed solar wind proton energy. Hatched regions indicate energies without data coverage. Lines show the median modelled flux of scattered (black dashed), sputtered (black dash-dotted), and total (solid red) negative hydrogen ions; the gray shading denotes the 68\% highest density interval of the total modelled flux.}
  \label{fig:NILS_surface_data}
\end{figure}

\begin{figure}[ht!]
\centering
\includegraphics[width=1\linewidth]{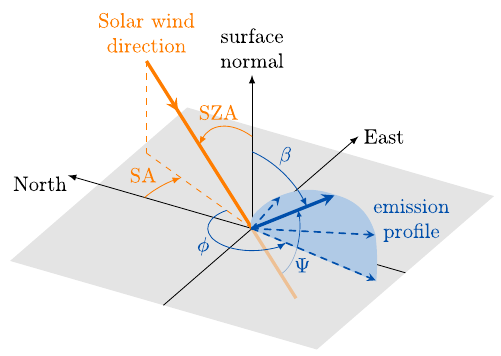}
\caption{Illustration of the solar wind impinging angles (orange) and emission angles (blue). An arbitrary angular emission profile is drawn, with dashed arrows representing possible emission directions. Both the Solar Zenith Angle (SZA) and the emission polar angle $\beta$ are defined relative to the surface normal. The Solar Azimuth (SA) angle is relative to the North direction, and the emission azimuthal angle $\phi$ is relative to SA. The total scattering angle $\Psi$ is the angle between the solar wind direction and the emission direction.}
\label{fig:all_angles}
\end{figure}

Figure~\ref{fig:NILS_surface_data} shows the differential number flux of negative hydrogen ions from the lunar surface for different observation geometries. The size of the vertical bars indicates the range of likely values in the (model-agnostic) estimates of the differential flux. The colour of the bars represents the significance of the observation. For energies below 100\,eV, the statistical uncertainty in the separation of hydrogen ions from electrons increases. The lines show the contributions from scattering and sputtering, as derived from the particle–surface interaction model described in the following section. While NILS recorded data, the average solar wind proton energy was $\overline{E}_\mathrm{sw}=466\,\mathrm{eV} \; (\pm19.6\,\mathrm{eV})$, with an average proton number density of $8.1\pm2.2\,\mathrm{cm}^{-3}$. The angular observation geometry is illustrated in Fig.~\ref{fig:all_angles}. The solar zenith angle ($\mathrm{SZA}$) varied between
52$^\circ$ and 47$^\circ$, the solar azimuth angle ($\mathrm{SA}$) varied from $43^\circ$E to $28^\circ$E and the average azimuthal emission angle was $\overline{\phi} = 213^\circ$.
The various panels in Fig.~\ref{fig:NILS_surface_data} show data for different intervals of the polar emission angle $\beta$ .


\section{Interaction model}
\label{sec:interaction_model}

\subsection{A general model}

We define the differential number flux $\mathcal{J}$ $[1/(\mathrm{cm}^2\,\mathrm{sr}\,\mathrm{eV}\,\mathrm{s})]$ as the flux of emitted particles from a given species, \emph{independent} of their charge state. We decompose $\mathcal{J}$ into a total flux $\mathcal{A} \; [1/(\mathrm{cm^2\,s} )]$ and two normalized functions ${\mathcal{J}_E} \; [1/\mathrm{eV}]$ and ${\mathcal{J}_\Omega} \; [1/\mathrm{sr}]$ that describe the energy and angular distribution of the flux

\begin{equation}
    \mathcal{J}\left(E,\Omega\right) = \mathcal{A} \cdot {\mathcal{J}_E}\left(E;\Omega\right) \cdot {\mathcal{J}_{\Omega}}\left(\Omega\right)\;,
    \label{eq:flux_components}
\end{equation}

with $\int{\mathcal{J}_E\left(E;\Omega\right)\mathrm{d}E} = 1$ and $
\int{\mathcal{J}_{\Omega}\left(\Omega\right)\mathrm{d}\Omega} = 1$, where $E\;[\mathrm{eV}]$ is the energy of the emitted particle, and $\Omega = \left(\beta, \phi\right)$ is the emission solid angle, with $\beta\;\left[\mathrm{rad}\right]$ and $\phi\;\left[\mathrm{rad}\right]$ the polar and  azimuthal emission angles, respectively (see Fig. \ref{fig:all_angles}). We factorize the total flux $\mathcal{A}$ into

\begin{equation}
    \mathcal{A} = \eta \cdot f_\perp\;,
    \label{eq:amplitude}
\end{equation}

with $f_\perp \; \left[1/\left(\mathrm{cm^2\,s}\right)\right]$ the incident particle flux \emph{normal to the emitting surface}. The yield $\eta$ defines the total number of particles emitted from the surface per incident particle. We rewrite Eq.~\ref{eq:flux_components} in expanded form as

\begin{equation}
    \mathcal{J} = f_\perp \cdot \mathcal{J}_E \cdot \left[\eta\mathcal{J}_{\Omega}\right]\;,
    \label{eq:general_flux_model}
\end{equation}

where $\left[\eta\mathcal{J}_{\Omega}\right]$ defines the angular yield, that is, the number of particles emitted into a given solid angle per incident particle, integrated over all emission energies. 

\subsection{Charge states of emitted particles}

The quantity $\mathcal{J}$ defines the differential flux of all emitted particles of a given species, that is, regardless of their charge state. To obtain the differential flux for a given charge state $q$, we introduce the probability of ionization $\mathrm{P}^q$ as

\begin{equation}
    \mathrm{P}^q \equiv \mathcal{J}^q / \mathcal{J}\;,
    \label{eq:charge_states_definition}
\end{equation}

where $\mathcal{J}^q$ is the differential flux of emitted particles of charge state $q$. We neglect multiply charged ions, and only consider positive ions $(q=+)$, negative ions $(q=-)$, and neutral particles $(q=0)$. The probability of ionization satisfies the following condition

\begin{equation}
    \mathrm{P}^+ + \mathrm{P}^- + \mathrm{P}^0 = 1\;.
\end{equation}

The probability of ionization depends on the species, energy, emission direction of the emitted particle, and the properties of the surface.

\subsection{Independence from the projectile charge state}

The definition of $\mathcal{J}$ is based on the postulate that the charge state of the projectile does not affect the collision dynamics leading to the eventual emission of a particle(s). In other words, the charge state of $f_\perp$ has no influence. For the scattering of hydrogen and oxygen atoms or ions in the hundreds of eV to keV range, the fraction of negative ions is independent of the initial charge state of the projectile \citep{Lienemann_2011}; a charge state equilibrium is achieved almost instantaneously during the interaction process. Auger neutralization dominates the charge-exchange at long range, whereas resonant charge transfer dominates at close range. The characteristic timescales of these charge-exchange processes are on the order of femtoseconds or less \citep{Wang_2001}, while a binary collision occurs hundreds or thousands of times slower (in the case of keV projectiles). Similarly, \citet{Schenkel_1997} and \citet{Pesic_2007} found that for highly charged ions (q=7--65) with keV energies, the first electron capture typically occurs about 60\,a.u. ($\approx 3$\,nm) above the surface, followed by complete neutralization within a few tens of femtoseconds.

\subsection{Negative ionization probability}
\label{subsec:negative_ioniz_prob}


After leaving the surface, a particle continues to exchange electrons with the surface until it reaches its final charge state at distances of several nanometres. The final charge state of particles emitted from a metallic surface is dominantly formed by resonant charge transfer \citep{Los_1990, Gainullin_2020}. Charge transfer occurs through electron tunnelling across the potential barrier between the metal conduction band and energetically-overlapping atomic states of the projectile.

The probability of negative ionization $\mathrm{P}^-$, defined as the probability that a particle reaches a negative charge state after leaving the surface, can be expressed for metallic surfaces as \citep[Eq. 13]{Eckstein_1981}

\begin{equation}
    \mathrm{P}^-\left(v_\perp\right) = \exp\left(-v_s/v_\perp\right) \left[1 - \exp\left(-v_f/v_\perp\right)\right]\;,
    \label{eq:negative_ionization_probability_metal}
\end{equation}

with $v_\perp = \cos\beta \sqrt{2E/m}$ the velocity component of the emitted particle along the surface normal in m/s, $m$ the mass of the particle in kg, $E$ the energy of the particle in J, and $(v_f, v_s)$ the characteristic velocities associated with the formation and survival of the negative charge state, respectively. 

\citet{Verbeek_1980} showed that for the emission of hydrogen atoms, the probability of negative ionization, $\mathrm{P}^-$, depends on the emission energy differently from the probability of positive ionization, $\mathrm{P}^+$. While $\mathrm{P}^+$ increases with energy, $\mathrm{P}^-$ reaches a maximum around 2–-3\,keV and then decreases. Additionally, they found that the probability of negative ionization increases as the surface work function decreases, consistent with the dominance of resonant charge transfer: a lower work function reduces the energy gap between the conduction band and the electronic states of the projectile, increasing the efficiency of resonant exchange over larger distances. For emitted particles with energies in the hundreds of eV, the probability of negative ionization can be simplified to \citep[Eq.~2-4]{Wucher2008}

\begin{equation}
    \mathrm{P}^-\left(v_\perp\right) \approx \mathrm{S}\exp\left(-v_c/v_\perp\right)\;,
    \label{eq:P_expo_dependece}
\end{equation}

with $v_c$ the critical velocity of the system and $\mathrm{S}$ a constant that controls the probability at high emission velocities ($v_\perp \gg v_c$). The same relation was found by \citet{Lang_1983} when describing the probability of negative ionization for atoms sputtered off a metal surface. 

\begin{figure}[t]
    \centering
    \includegraphics[width=1\linewidth]{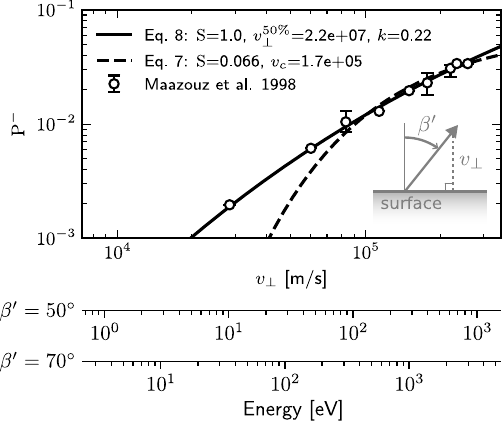}
    \caption{Probability of negative ionization as a function of the perpendicular emission velocity, $v_\perp$, for 1\,keV ($\approx 4.37\times 10^{5}\,\mathrm{m/s}$) hydrogen ions scattering off silicon. The mapping to the emission energy for different microscopic emission angles $\beta{'}$ (Section \ref{sec:microscaleabeta}) is shown below the figure. Data points (open circles) are taken from \citet[Fig.~5a]{Maazouz_1998}, with the best fit from Eq.~\ref{eq:P_expo_dependece} shown as a dashed line and from Eq.~\ref{eq:negative_ionization_probability_silicon} as a solid line.}
    \label{fig:negative_ionization_probability}
\end{figure}

In contrast to conductive surfaces, the lunar regolith consists of insulating materials with high work functions and large bandgaps. Therefore, the quasi-free electron model used for metal surfaces is not applicable. Nevertheless, \citet{Borisov_2000} showed that despite these unfavourable conditions for resonant charge exchange, insulating surfaces can still be efficient producers of negative ions. For example, \citet{Wieser_2002} and  \citet{Wurz2006} showed that about 3-7\% of protons scattering of a $\mathrm{MgO}$ surface at grazing incidence are converted to negative hydrogen. 

Figure~\ref{fig:negative_ionization_probability} shows measurements of the probability of negative ionization for 0.5-–4\,keV hydrogen ions scattering from semiconducting silicon, as reported by \citet{Maazouz_1998}. The model for metallic surfaces (Eq.~\ref{eq:P_expo_dependece}) provides a reasonable fit at large $v_\perp$ but underestimates the negative ionization probability at small $v_\perp$. As in the metallic case, the probability of negative ionization on semiconductors or insulators depends on the component of the ion velocity perpendicular to the surface. We describe this dependence using a functional form similar to that in \citet[Eq.~11]{Demkov1963}:

\begin{equation}
\mathrm{P}^-(v_\perp) = \mathrm{S} \,\mathrm{sech}^2\left[\mathrm{arcosh}(\sqrt{2}) \left( \dfrac{v_\perp^{50\%}}{v_\perp}\right)^k\right]\;,
\label{eq:negative_ionization_probability_silicon}
\end{equation}

where the parameter $v_\perp^{50\%}$ is the perpendicular speed for which $\mathrm{P}^-(v_\perp^{50\%})$ is half of its maximum value, that is, S/2; $k$ controls the steepness of the function; $\mathrm{S}$ is the ionization probability for $v_\perp\rightarrow\infty$. The values $\mathrm{S}=1$, $v_\perp^{50\%} = 6.4\times10^6$ m/s and $k=0.26$ give a good fit to the data in Fig.~\ref{fig:negative_ionization_probability} over a wide range of $v_\perp$. While $\mathrm{S}=1$ is not physically valid, the model holds for the range of perpendicular velocities relevant to this study. 

The atomic composition of the surface also plays a role: the presence of oxygen atoms is known to increase the probability of positive and, to a lesser extent, negative ionization \citep{Wucher2008, Wucher2013}.

\subsection{Emission angles at microscopic scales}
\label{sec:microscaleabeta}

The emission angle, $\beta$, is defined as the polar angle between the surface normal and the emission direction as shown in Fig.~\ref{fig:all_angles}. However, the definition of the surface normal depends on the spatial scale considered. When parametrizing the angular yield from in-situ data, we use the macroscopic (meter-scale) surface normal (see Sect.~\ref{subsec:lunar_topography}) so that our results can be compared with studies based on orbital observations. However, the establishment of the final charge state of an emitted particle is a process that occurs only nanometres above the surface. For that process, the emission angle must therefore be defined with respect to the microscopic surface normal and not the macroscopic average. We denote this microscopic emission angle by $\beta'$.

We use the theory of \citet{Szabo_2022a} to construct a mapping between the macroscopic angle $\beta$ and the microscopic angle $\beta'$. For the lunar regolith, the average microscopic emission angle $\overline{\beta'}$ [rad] corresponding to a given macroscopic angle $\beta$ [rad] is well approximated by the third-order polynomial (a more detailed discussion is provided in Appendix~\ref{appendix:regolith_rough_surface})

\begin{equation}
\overline{\beta'}\left(\beta\right) = -0.33\,\beta^3 + 0.81\,\beta^2 + 0.039\,\beta + 0.48\;.
\label{eq:average_polar_emission_angle_mapping}
\end{equation}

The NILS instrument covers the macroscopic polar angles $\beta\in\left[45,\,90\right]$ deg, corresponding to the average microscopic angle $\overline{\beta'}\in\left[49,\,72\right]$ deg. Due to the high roughness of the lunar regolith, a particle emitted at a grazing angle ($\beta' \gtrsim 70$\,deg) will likely impact the surface again and is therefore not measured by NILS.

\subsection{Putting it all together}

We first define the perpendicular influx $f_\perp$ as

\begin{equation}
    f_\perp = f_\mathrm{in} \cos\left(\mathrm{SZA}\right)\;,
    \label{eq:f_perp_f_in}
\end{equation}

where $f_\mathrm{in} \; [1/(\mathrm{cm}^2\;\mathrm{s})]$ is flux of the incident particles and $\mathrm{SZA}$ the angle between the incident direction and the surface normal, as schematized in Fig.~\ref{fig:all_angles}. The actual solar wind incident direction may slightly differ from the Sun direction due to aberration, but we use the term Solar Zenith Angle (SZA) for notational convenience. The formalism naturally applies to any directed incident particle population. 

We now assume that an emitted particle is either sputtered from the surface or originates from the scattering of precipitating particles. We express the differential number flux $\mathcal{J}$ as a sum of the two processes: 

\begin{equation}
    \mathcal{J} = \mathcal{J}^{\mathrm{sc}} + \mathcal{J}^{\mathrm{sp}}\;, 
    \label{eq:sc_sp_sum}
\end{equation}

where $\mathcal{J}^{\mathrm{sc}}$ denotes the scattered flux and $\mathcal{J}^{\mathrm{sp}}$ the sputtered flux. Combining Eqs. \ref{eq:general_flux_model}, \ref{eq:charge_states_definition}, \ref{eq:f_perp_f_in} and \ref{eq:sc_sp_sum}, we obtain the differential number flux of a specific charge state $q$

\begin{equation}
\begin{split}
    \mathcal{J}^q\left(E,\Omega\right) =& \mathrm{P}^q f_{\mathrm{in}} \cos\left(\mathrm{SZA}\right)\;\cdot\\
    &\big\{\\
    &\quad\mathcal{J}_E^\mathrm{sc}\left(E; \Omega\right) \; \left[\eta^\mathrm{sc} \; \mathcal{J}_{\Omega}^\mathrm{sc}\left(\Omega\right)\right]\\
    &+\\
    &\quad\mathcal{J}_E^\mathrm{sp}\left(E; \Omega\right)\; \left[\eta^\mathrm{sp} \; \mathcal{J}_{\Omega}^\mathrm{sp}\left(\Omega\right)\right] \\
    &\big\}\;,
    \label{eq:grand_model}
\end{split}
\end{equation}

with $\eta^\mathrm{sc} \in \left[0,1\right]$ the scattering yield, which naturally cannot exceed one. The sputtering yield can, in principle, exceed one with $\eta^\mathrm{sp}\ge0$. The energy distribution of the scattered and sputtered particles are modelled in the following sections.

\subsection{The sputtered component}
\label{subsec:sputtering}


\begin{figure}[ht!]
    \centering
    \includegraphics[width=1\linewidth]{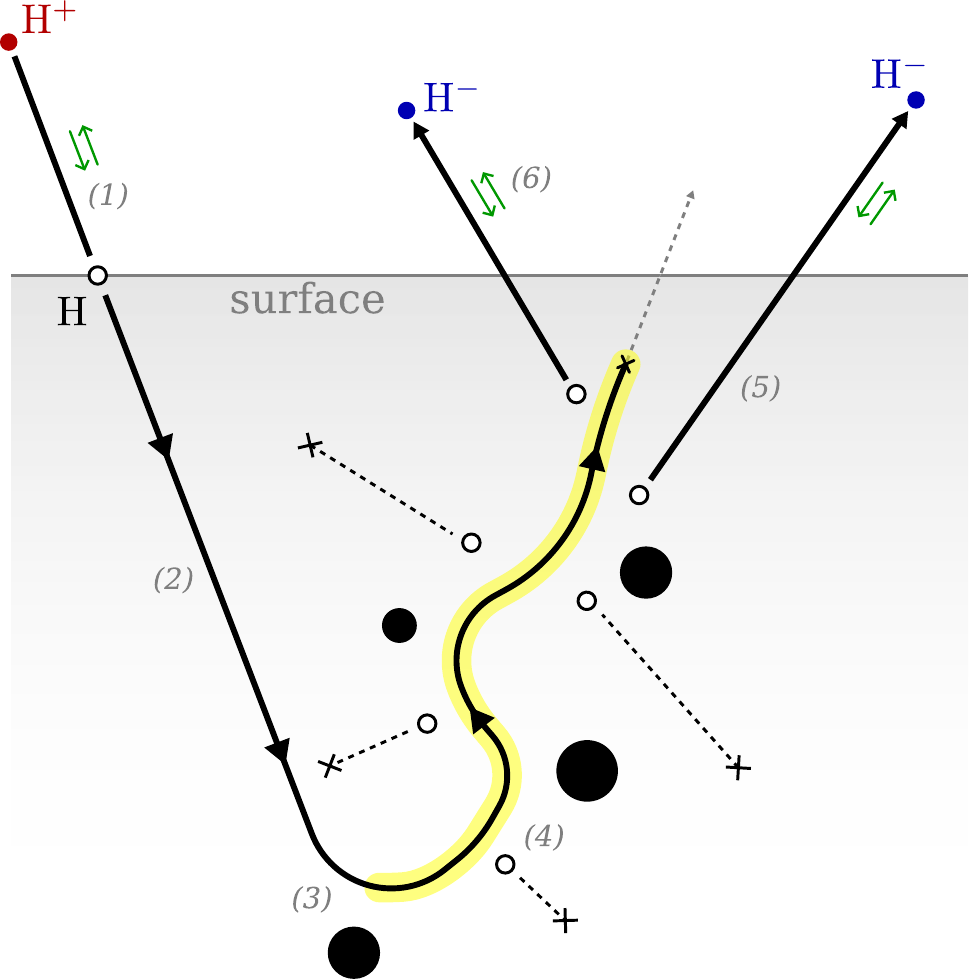}
    \caption{Schematized sputtering induced by light ions. A proton (filled red circle) directed toward the surface (1) rapidly neutralizes to an hydrogen atom (open circle); (2) penetrates the material and loses a fraction $\gamma_\mathrm{extra}$ of its energy; (3) undergoes a large-angle scattering on a surface atom (filled black circles); (4) creates an isotropic emission of hydrogen primary knock-on atoms (PKAs, open circles) while losing additional energy inelastically (yellow overlay); most knock-on hydrogen atoms do not escape the surface (crosses) while those near the surface (5) have an increased probability of escaping the surface; (6) charge-exchange transfers (green arrows) may ultimately lead to a negative charge state (filled blue circle).}
    \label{fig:sputtering_schematic}
\end{figure}

The energy distribution of atoms sputtered from the lunar surface under solar wind irradiation is described by the analytical model of \cite{Ono_2005}. Their model extends the earlier formulation of \cite{KENMOTSU_2004}, which was successfully applied by \citet{Wieser_2024} to describe the sputtering of hydrogen atoms from the lunar surface. Unlike the original model by \citet{KENMOTSU_2004}, the extension accounts for both elastic and inelastic energy losses, the latter being important in the hundreds of eV to few keV range. For projectile energies of hundreds of eV, the extended model shows better agreement with simulations than the earlier model \citep[Figs. 2 and 3]{Ono_2005}.

Both models are based on the same sputtering mechanism described by \citet{KENMOTSU_2004}. They consider a projectile that penetrates a solid surface, backscatters at a large angle from a surface atom, and loses energy through inelastic and elastic collisions--the latter generating \emph{recoil} atoms \citep{Sigmund1969}, which are surface atoms displaced from their original lattice position. Recoil atoms with sufficient energy to exit the surface are said to be sputtered. Atoms directly struck by the projectile are referred to as primary recoil atoms, or Primary Knock-on Atoms (PKAs). For light projectiles (for example, hydrogen, deuterium, helium), secondary, tertiary, and higher-order recoil atoms rarely contribute significantly to sputtering, so the models primarily focus on PKAs. Figure \ref{fig:sputtering_schematic} illustrates the different stages of the sputtering process as considered in this section.

While the original model by \citet{Ono_2005} describes the sputtering of a heavy \emph{single}-component amorphous material by light ions, we modified it to describe the sputtering of hydrogen by light ions impacting a heavy \emph{multi}-component amorphous material. We first introduce the notation $(\mathrm{P}\rightarrow\mathrm{S})$ to indicate that a projectile atom $\mathrm{P}$ is interacting with an arbitrary surface atom of species $\mathrm{S}$. To emphasize the special case of a projectile atom $\mathrm{P}$ interacting with a hydrogen atom in the material, we use the notation $(\mathrm{P}\rightarrow\mathrm{H})$. 


We rewrite Eq.\,1 from \citet{Ono_2005}--which describes the energy loss of a projectile as it traverses the surface and the recoiling of hydrogen atoms--to account for the multi-species nature of the interaction:

\begin{equation}
\begin{split}
\overbrace{r_\mathrm{H}\dfrac{\mathrm{d}\sigma_{\left(\mathrm{P}\rightarrow \mathrm{H}\right)}\left(E,E_0\right)}{\mathrm{d}E_0}}^{\left(1\right)}&= \overbrace{\dfrac{\partial F_{\mathrm{PKA}}\left(E,E_0\right)}{\partial E}}^{\left(2\right)} \\
&\cdot\sum_{\mathrm{S}} r_{\mathrm{S}} \left[ \underbrace{S_e^{\left(\mathrm{P}\rightarrow \mathrm{S}\right)}\left(E\right)}_{\left(3\right)} + \underbrace{\int_0^{T_{\max}^{\left(\mathrm{P}\rightarrow \mathrm{S}\right)}} T \mathrm{d}\sigma_{\left(\mathrm{P}\rightarrow \mathrm{S}\right)}\left(E,T\right)}_{\left(4\right)} \right]\;,
\end{split}
\label{eq:sputtering_theory}
\end{equation}

where $\mathrm{d}\sigma_{\left(\mathrm{P}\rightarrow \mathrm{S}\right)}$ [nm$^2$] is the differential elastic scattering cross section for a projectile $\mathrm{P}$ interacting with a surface atom of species $\mathrm{S}$; similarly, $\mathrm{d}\sigma_{\left(\mathrm{P}\rightarrow \mathrm{H}\right)}$ is the differential elastic scattering cross section for a projectile $\mathrm{P}$ interacting with a surface hydrogen atom; $F_\mathrm{PKA}$ [1/eV] is the energy distribution of hydrogen PKAs produced by the projectile; $E$ [eV] is the energy of the projectile; $E_0$ [eV] is the initial energy of a hydrogen PKA; $S_e^{\left(\mathrm{P}\rightarrow \mathrm{S}\right)}$ [eV\,nm$^2$] is the electronic stopping cross section for a projectile P interacting with the electronic cloud of a surface atom of species S; $r_\mathrm{S}$ is the relative atomic concentration of species $\mathrm{S}$ in the material; similarly, $r_\mathrm{H}$ is the relative atomic concentration of hydrogen atoms in the material, $T$ [eV] is the recoil energy of surface atoms; and $T_{\max}^{\left(\mathrm{P}\rightarrow \mathrm{S}\right)}$ [eV] is the maximum energy of recoil atoms of species $\mathrm{S}$ after an elastic collision with a projectile $\mathrm{P}$. The various source and sink terms in Eq.~\ref{eq:sputtering_theory} are:

\begin{itemize}
    \item[(1)] Source term, that is, the number of hydrogen PKAs with an energy $E_0$ created by a projectile with energy $E$.
    \item[(2)] The rate of change in the number of hydrogen PKAs of energy $E_0$ that the projectile can still produce as its energy decreases.
    \item[(3)] The energy loss rate of the projectile due to inelastic collisions.
    \item[(4)] The energy loss rate of the projectile due to elastic collisions.
\end{itemize}

\subsubsection{Initial backscattering}

The maximum energy of a recoil atom of species $\mathrm{S}$ after an elastic collision with a projectile $\mathrm{P}$ is defined as

\begin{subequations}
\begin{align}
    T_{\max}^{\left(\mathrm{P}\rightarrow \mathrm{S}\right)} &= E_\mathrm{in} \Gamma_{\left(\mathrm{P}\rightarrow \mathrm{S}\right)}\;,\\
    \Gamma_{\left(\mathrm{P}\rightarrow \mathrm{S}\right)} &= \underbrace{\left(1-\gamma_{\mathrm{back}}\right) \left(1-\gamma_\mathrm{extra}\right)}_{\left(1\right)} \;\overbrace{\gamma_{\left(\mathrm{P}\rightarrow \mathrm{S}\right)}}^{\left(2\right)}\;,
\end{align}
\label{eq:T_max}
\end{subequations}

where $E_\mathrm{in} \; [\mathrm{eV}]$ is the initial energy of the projectile, and $\gamma_{\left(\mathrm{P}\rightarrow \mathrm{S}\right)}$ defines the maximum fraction of energy transferred from the projectile $\mathrm{P}$ to a target atom $\mathrm{S}$ during a binary elastic collision, as given by \citep[Eq. 8]{Niehus_1993}

\begin{equation}
\gamma_{\left(\mathrm{P}\rightarrow\mathrm{S}\right)}=\dfrac{4M_{\mathrm{P}}M_{\mathrm{S}}}{\left(M_{\mathrm{P}}+M_{\mathrm{S}}\right)^2}\;,
\label{eq:gamma}
\end{equation}

where the mass of the projectile is indicated by $M_\mathrm{P}$ and the mass of the surface atom by $M_\mathrm{S}$. In the special case of hydrogen–-hydrogen interactions, all energy is transferred, that is, $\gamma_{\left(\mathrm{H}\rightarrow \mathrm{H}\right)} = 1$. The energy loss factor $\gamma_\mathrm{back}$ defines the energy that the projectile looses when backscattering at large-angle from a surface atom \citep["backscattered near 180° by target atoms"]{Ono_2005}. For simplicity, we approximate this factor as $\gamma_\mathrm{back} \approx \sum_\mathrm{S}r_\mathrm{S} \gamma_{\left(\mathrm{P}\rightarrow\mathrm{S}\right)} = 0.21$ (see Table.~\ref{tab:elemental_composition}).

An additional empirical energy loss factor, $\gamma_{\mathrm{extra}}$, is introduced to increase the flexibility of the model, accounting for effects not explicitly included, such as the energy lost by the incident particle prior to large-angle backscattering and the energy lost by hydrogen PKAs as they traverse the material up to the surface. The factor $\Gamma_{\left(\mathrm{P}\rightarrow \mathrm{S}\right)}$ thus quantifies the \emph{maximum} and \emph{total} energy transfer between the projectile and a surface atom. Equation~\ref{eq:T_max} conceptually separates the transport of the projectile through the material into two stages:

\begin{itemize}
    \item[(1)] the initial large-angle backscattering of the projectile $\mathrm{P}$ by a surface atom $\mathrm{S}$ (step 3 in Fig. ~\ref{fig:sputtering_schematic}), and 
    \item[(2)] the subsequent elastic binary collisions with surface atoms (step 4 in Fig. ~\ref{fig:sputtering_schematic}).
\end{itemize}

\subsubsection{Elastic losses}

The differential scattering cross section $\mathrm{d}\sigma_{\left(\mathrm{P}\rightarrow \mathrm{S}\right)}$ is approximated by a  parametrized expression \citep[Eq.~7]{Ono_2005}, which parameters depend on the scattering regime that dominates the interaction. To determine the scattering regime, we evaluate $R=\epsilon_{\left(\mathrm{P}\rightarrow\mathrm{S}\right)}\left(E\right)\sqrt{T/T_{\mathrm{max}} ^{\left(\mathrm{P}\rightarrow \mathrm{S}\right)}}$ at the maximum transferred energy $T = T_{\max}$. 
The reduced energy $\epsilon$ is defined in as \citet[Eq. 5]{Ono_2005}

\begin{equation}
    \epsilon_{\left(\mathrm{P}\rightarrow \mathrm{S}\right)}\left(E\right) = \dfrac{4\pi \epsilon_0}{e^2}\dfrac{a^{\left(\mathrm{P},\;\mathrm{S}\right)}}{Z_\mathrm{P}Z_\mathrm{S}} \dfrac{M_\mathrm{S}E}{M_\mathrm{P} + M_\mathrm{S}}\;,
    \label{eq:reduced_energy}
\end{equation}

with $Z_{\mathrm{P}}$ the atomic number of the projectile, $Z_{\mathrm{S}}$ the atomic number of the surface atom, $\epsilon_0$ the vacuum permittivity, $e$ is the elementary charge, and $a^{\left(\mathrm{P},\;\mathrm{S}\right)}$ the Thomas-Fermi screening length defined as

\begin{equation}
    a^{\left(\mathrm{P},\;\mathrm{S}\right)}=4.685 \times10^{-2} \left(Z_{\mathrm{P}}^{2/3}+Z_{\mathrm{S}}^{2/3}\right)^{-1/2} \; \left[\mathrm{nm}\right].
\end{equation}

The reduced energy is calculated using the energy of the backscattered ion, $E_{\mathrm{back}} \approx E_{\mathrm{in}}\left(1-\gamma_\mathrm{back}\right)$, prioritizing high-energy transfer collisions as these are most likely to produce PKAs with sufficient energy to contribute significantly to sputtering. For solar wind protons interacting with lunar regolith, the value of $R$ is in the range of $1.18\times 10^{-2} \le R \leq 1.11$. This corresponds to the scattering regime identified as \textit{Region II} in the model of \citet{Ono_2005}, with the differential scattering cross section expressed as

\begin{equation}
\mathrm{d}\sigma_{\left(\mathrm{P}\rightarrow \mathrm{S}\right)}\left(E,T\right) = C_{\left(\mathrm{P}\rightarrow \mathrm{S}\right)} E^{-1/2} T^{-3/2}\mathrm{d}T\,,
\label{eq:diff_elatic_crosssection}
\end{equation}

with $C_{\left(\mathrm{P}\rightarrow \mathrm{S}\right)} \; \left[\mathrm{eV}\;\mathrm{nm}^2\right]$ a constant controlling the elastic scattering cross section defined as 

\begin{equation}
    C_{\left(\mathrm{P}\rightarrow \mathrm{S}\right)} = 0.276\pi a^{\left(\mathrm{P},\;\mathrm{S}\right)}Z_{\mathrm{P}}Z_{\mathrm{S}}\sqrt{\dfrac{M_{\mathrm{P}}}{M_{\mathrm{S}}}} \dfrac{e^2}{{4\pi\epsilon_0}}\;,
    \label{eq:elastic_cross_section}
\end{equation}

where, for convenience, we can approximate the constant $e^2/4\pi\epsilon_0 \approx 1.44 \;\left[\mathrm{eV}\;\mathrm{nm}\right]$.

\subsubsection{Inelastic losses}
\label{subsubsec:inelastic_losses_general}

Solar wind protons have energies from a few hundred eV to a few keV. Their interaction with surfaces falls within the Lindhard–Scharff regime \citep{Lindhard1961}, where the electronic stopping power $S_e^{\left(\mathrm{P}\rightarrow \mathrm{S}\right)}$ is approximately proportional to the projectile velocity:

\begin{subequations}
\begin{align}
    S_e^{\left(\mathrm{P}\rightarrow \mathrm{S}\right)}\left(E\right) &= K_{\left(\mathrm{P}\rightarrow \mathrm{S}\right)} \sqrt{E}\;\text{ , with}\\
    K_{\left(\mathrm{P}\rightarrow \mathrm{S}\right)} &= \dfrac{1.216\times 10^{-2} \;Z_{\mathrm{P}}^{7/6} \;Z_{\mathrm{S}}}{\sqrt{M_{\mathrm{P}}} \; \left(Z_{\mathrm{P}}^{2/3}+Z_{\mathrm{S}}^{2/3}\right)^{3/2}} \; \left[ \mathrm{eV}^{1/2} \; \mathrm{nm}^2\right]\;.
\end{align}    
\label{eq:electronic_stopping_power}
\end{subequations}

For clarity, we write as $K_{\left(\mathrm{P}\right)} = \sum_\mathrm{S} r_\mathrm{S} K_{\left(\mathrm{P}\rightarrow \mathrm{S}\right)}$ the inelastic energy loss constant averaged over the elemental composition of the material. Inelastic losses are indicated by the yellow overlay in Fig. ~\ref{fig:sputtering_schematic}.

\subsubsection{Recoil density and energy distribution}

After some algebraic rearrangement of Eq.~\ref{eq:sputtering_theory}, we obtain the hydrogen PKA density within the material

\begin{equation}
    \dfrac{\partial F_{\mathrm{PKA}}\left(E,E_0\right)}{\partial E} = \dfrac{r_\mathrm{H}C_{\left(\mathrm{P}\rightarrow \mathrm{H}\right)}E_0^{-3/2}E^{-1/2}}{K_{\left(\mathrm{P}\right)} \sqrt{E} + 2\sum_{\mathrm{S}} r_{\mathrm{S}} C_{\left(\mathrm{P}\rightarrow \mathrm{S}\right)}\sqrt{\Gamma_{\left(\mathrm{P}\rightarrow\mathrm{S}\right)}}}\;.
    \label{eq:PKA_density}
\end{equation}

A projectile of energy $E$ can create an hydrogen PKA of energy $E_0$ if it has sufficient energy, that is, $E \ge E_0 / \Gamma_{\left(\mathrm{P}\rightarrow\mathrm{H}\right)} = E_{\min}$. We integrate Eq.~\ref{eq:PKA_density} over $E$ from $E_{\min}$ to the initial projectile energy $E_\mathrm{in}$, giving

\begin{subequations}
\begin{align}
    F_{\mathrm{PKA}}\left(E_0; E_\mathrm{in}\right) &\propto E_0^{-3/2} \ln \left[\dfrac{\sqrt{E_\mathrm{in}} + \mathcal{B}_{\left(\mathrm{P}\right)}}{\sqrt{\dfrac{E_0}{\Gamma_{\left(\mathrm{P}\rightarrow\mathrm{H}\right)}}} + \mathcal{B}_{\left(\mathrm{P}\right)}}\right]\;,\\
    \mathcal{B}_{\left(\mathrm{P}\right)} &= 2\dfrac{\sum_{\mathrm{S}} r_{\mathrm{S}} C_{\left(\mathrm{P}\rightarrow \mathrm{S}\right)}\sqrt{\Gamma_{\left(\mathrm{P}\rightarrow\mathrm{S}\right)}}}{K_{\left(\mathrm{P}\right)}}\;\left[\mathrm{eV}^{1/2}\right]\,.
\end{align}
    \label{eq:integrated_PKA_density}
\end{subequations}

We write $F_{\mathrm{PKA}}$ up to a multiplicative constant that is energy-independent and therefore does not affect the energy distribution of hydrogen PKAs. The energy distribution of hydrogen PKAs that escape the surface and are sputtered is obtained under the following assumptions: (i) the probability of a hydrogen PKA escaping the surface decreases exponentially with its creation depth, and (ii) the hydrogen PKAs are created isotropically very close to the surface. Hydrogen PKAs need to have enough energy to overcome the surface binding energy $U$ [eV] and escape the surface. We integrate Eq.~\ref{eq:integrated_PKA_density} over $E_0$ from $E_0=U$ to $E_0=T_{\max}^{\left(\mathrm{P}\rightarrow \mathrm{H}\right)}$, and over the creation depth and solid angles to obtain an expression for the energy distribution of sputtered hydrogen atoms when a projectile $\mathrm{P}$ irradiates a multi-component surface:

\begin{equation}
    \boxed{\mathcal{J}_E^{\mathrm{sp}}\left(E;E_\mathrm{in}, U\right) = \dfrac{E}{\mathtt{n}_{\left(\mathrm{P}\right)}\left(E+U\right)^{5/2}} \ln\left(\dfrac{\sqrt{E_{\mathrm{in}}} + \mathcal{B}_{\left(\mathrm{P}\right)}}{\sqrt{\dfrac{E+U}{\Gamma_{\left(\mathrm{P}\rightarrow\mathrm{H}\right)}}} + \mathcal{B}_{\left(\mathrm{P}\right)}} \right)} \; ,
    \label{eq:sputtering_model_final}
\end{equation}

for $E\in\left[0, E_\mathrm{in} \Gamma_{\left(\mathrm{P}\rightarrow\mathrm{H}\right)} - U\right]$, and 0 otherwise, where $\mathtt{n}_{\left(\mathrm{P}\right)}$ is a normalization factor such that $\int_0^\infty \mathcal{J}_E^\mathrm{sp}(E)\mathrm{d}E=1$.

An estimation for $\mathcal{B}$ can be obtained using averages of the elemental composition model given in Table~\ref{tab:elemental_composition}. Figure \ref{fig:ono_model_example} shows examples of the energy distributions of hydrogen atoms sputtered from the lunar regolith for different values of $\gamma_{\mathrm{extra}}$.

\subsubsection{Dependence on $\mathcal{B}$}

The constant $\mathcal{B}$ is analytically complex as it depends on many parameters. However, for estimating the sputtering probability across the entire energy spectrum, $\mathcal{B}$ can usually be neglected ($\mathcal{B}=0$) without significant loss of accuracy.

Let $a\equiv \sqrt{E_{\mathrm{in}}}$ and $c\equiv \sqrt{\left(E+U\right)/\Gamma_{\left(\mathrm{P}\rightarrow\mathrm{H}\right)}}$. For $\mathcal{B} \ll a,c$, a first-order Taylor series expansion of the logarithm in Eq.~\ref{eq:sputtering_model_final} about $\mathcal{B}=0$ gives $\ln\left(a/c\right) + \mathcal{B}\left(1/a - 1/c\right) + \mathcal{O}\left(\mathcal{B}^2\right)$. For $a\approx c$, the first-order derivative is 0, making the logarithmic term in Eq. \ref{eq:sputtering_model_final}, and consequently $\mathcal{J}_E^{\mathrm{sp}}$, independent of $\mathcal{B}$. Even for $\mathcal{B} \not\ll a,c$, that is, when $\mathcal{B} \approx \sqrt{E_\mathrm{in}}$ as it is the case for solar wind energies, the first-order derivative term remains small if $a\approx c$. Therefore, the dependence on $\mathcal{B}$ can be still be neglected. The constant $\mathcal{B}$ starts to matter when $c \not= a$, that is $E \ll E_\mathrm{in}$. The high-energy end of the spectrum, however, is primarily controlled by energy of the projectile and the energy it loses while backscattering at large angles within the material.

\subsubsection{Alpha-induced sputtering efficiency}
\label{subsubsec:alpha_sputtering}

The solar wind is composed primarily of protons and alpha particles, both of which can sputter surface hydrogen atoms, although with different efficiencies. From \citet[Eqs.~14-II and 15-II][]{Ono_2005}, we find that the sputtering efficiency $\mathcal{Y}_{\left(\mathrm{P}\rightarrow \mathrm{H}\right)}$, defined as the number of sputtered hydrogen atoms (of any energy) per incident projectile $\mathrm{P}$, is equal to:

\begin{equation}
    \mathcal{Y}_{\left(\mathrm{P}\rightarrow \mathrm{H}\right)} \equiv \dfrac{\mathtt{n}_{\left(\mathrm{P}\right)}\,L\, C_{\left(\mathrm{P}\rightarrow \mathrm{H}\right)}}{K_{\left(\mathrm{P}\right)}}\;,
    \label{eq:sputtering_eff}
\end{equation}

with $L$ the collision mean free path of the hydrogen PKAs, and $\mathtt{n}$ the normalization factor introduced in Eq.~\ref{eq:sputtering_model_final}. The equation above has a direct physical interpretation: a larger scattering cross section ($C$) leads to more efficient PKA production, but this is counterbalanced by the ability of the projectile to lose energy via electronic stopping ($K$). The larger the mean free path, the larger the sputtered yield. Therefore, using Eq.~\ref{eq:sputtering_eff}, we find that, for the lunar regolith, an alpha particle sputters hydrogen atoms about $\mathcal{Y}_{\left(\mathrm{He}^{++}\rightarrow \mathrm{H}\right)} / \mathcal{Y}_{\left(\mathrm{H}^{+}\rightarrow \mathrm{H}\right)} \approx 2.8$ times more efficiently than a proton. The effect of $L$ cancels out because, regardless of the projectile, the sputtered atoms are always hydrogen.

Simulations of solar wind–ion–induced sputtering of various minerals, along with experimental measurements on lunar regolith samples, indicate that alpha-induced sputtering yields are approximately ten times higher than those produced by protons \citep{Morrissey2024, Broetzner2025}, exceeding our simplified estimate. We nevertheless keep our lower estimate for sake of consistency with the model presented here.

\begin{figure}[ht!]
    \centering
    \includegraphics[width=1\linewidth]{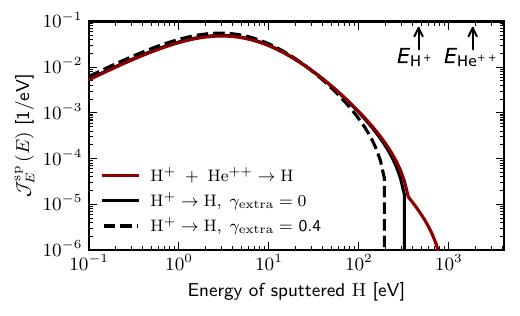}
    \caption{Energy distribution of hydrogen atoms sputtered by 300 km/s solar wind $\mathrm{H^+}$ and $\mathrm{He^{++}}$ ions impacting the lunar regolith (the energies of the projectiles are marked by an upward arrow). A surface binding energy of $U = 5\;\mathrm{eV}$ is assumed. Three models are shown: proton-induced sputtering without extra energy loss (solid line); proton-induced sputtering with extra energy loss (dashed line); and sputtering from both protons and alpha particles, assuming an alpha-to-proton ratio of 4\% and no extra energy loss (red solid line).}
    \label{fig:ono_model_example}
\end{figure}

Although alpha particles are more energetic and more efficient in sputtering hydrogen, their contribution to the energy distribution of sputtered hydrogen atoms is still small (see Fig.~\ref{fig:ono_model_example}) due to their low abundance in the solar wind. However, we include alpha particles when computing the total sputtering flux. During the NILS mission, the average solar wind alpha-to-proton number density ratio was approximately 4\%, based on the OMNI-2 database \citep{King_2005}. If alpha particles sputter hydrogen 2.8 times more efficiently than protons, their effect can be included as an effective increase in the precipitating proton flux: $\left(1 + k_\alpha\right) f_\mathrm{in}$, with the correction factor $k_\alpha = 2.8 \times 0.04 = 0.11$.

\subsection{The scattered component}
\label{subsec:scattering}

\begin{figure}[ht!]
    \centering
    \includegraphics[width=1\linewidth]{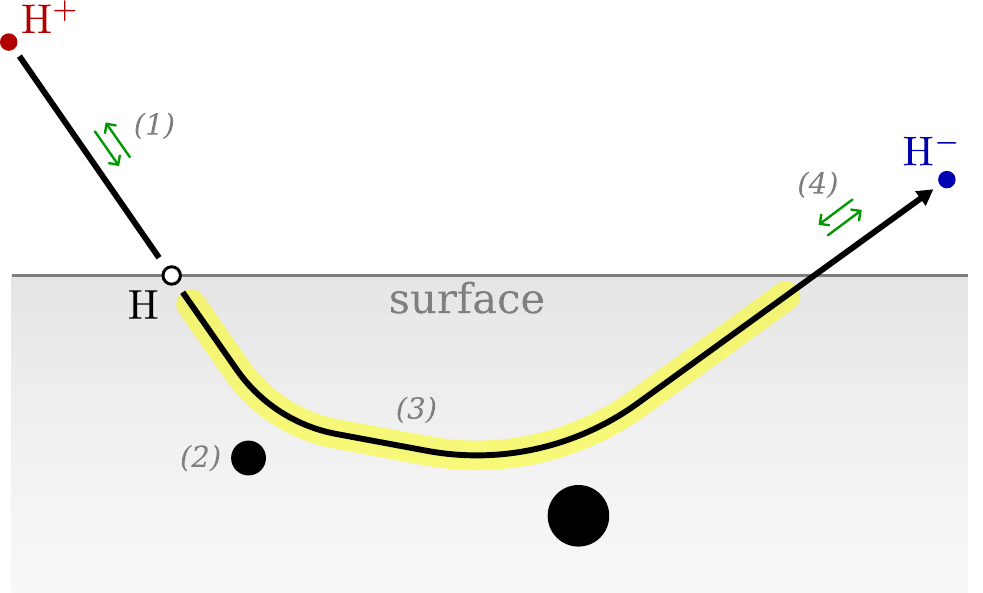}
    \caption{Schematized scattering of light ions from a surface. A proton (red filled circle) directed toward the surface (1) rapidly neutralizes; (2) scatter at small-to-medium angles off at least two surface atoms (black filled circles); (3) loses energy inelastically (yellow overlay); (4) exits the surface while charge-exchange transfers (green arrows) may lead to a final negative charge state (blue filled circle).}
    \label{fig:scattering_schematic}
\end{figure}

The energy distribution of light ions scattered from heavy materials has been extensively studied, particularly in the context of plasma–wall interactions in nuclear fusion reactors. At projectile energies of a few keV or lower, scattering occurs under two main regimes: single-collisional, dominating the high-energy end of the spectrum, and multiple-collisional, dominating the low-energy end. \citet{Tolmachev_1999} modelled light-ion scattering using a single-collision approach that includes both elastic and inelastic losses. Their model is versatile, as it applies to any ratio of the mean free path between elastic collisions to the total inelastic range.

In the energy region where multiple-collisions dominate, most theoretical models are derived by approximating the Boltzmann diffusion equation using various methods. \citet{Littmark1978} addressed light-ion scattering by approximating the ion depth distribution using the method of moments. A particular result is that neglecting the skewness of the depth distribution yields a Gaussian approximation of the energy distribution of scattered particles. An alternative study from \citet{Forlano_1996} used the discrete-streams method, assuming a power-law scattering cross section and neglecting inelastic losses. Their model successfully predicts the characteristic double-peak energy distribution observed for light ions scattered from heavy targets: the higher-energy peak corresponds to particles that undergo only a few collisions, whereas the lower-energy peak arises from the multiple-collision regime \citep{Broetzner2025a}. A later study by \citet{Falcone_1999} included inelastic losses but neglected elastic ones, an assumption valid for light-ions scattered from heavy targets \citep{Berger2005ESTAR}. This model shows good agreement with simulations, except for scattered ions with energies close to the incident energy, where the authors argue that adding elastic losses and fluctuations in inelastic losses would likely improve the agreement. \citet{Sukhomlinov1997} treated the elastic losses in a perturbative expansion to linearize the Boltzmann diffusion equation. Their model considers a realistic multi-component, inhomogeneous material, and they showed that increasing density with depth reduces backscattering of high-energy ions (deeper penetration, multiple scattering) but increases it for low-energy ions (shallow penetration, single scattering).

More recently, \citet{Afanas’ev_2025} showed that the theory of electron scattering is applicable for describing the scattering of light ions, and shows a particular good agreement with simulations for  projectiles with energies in the keV range and above. 

After considering all aforementioned analytical models, we chose the model of \citet{Forlano_1996} for its simplicity and adaptability for fitting purposes. Figure \ref{fig:scattering_schematic} summarizes the different stages of the scattering process considered in our model. 

\subsubsection{Adding inelastic losses}
\label{adding_inelastic_losses}

The model of \citet{Forlano_1996} considers ions that impact a surface at normal incidence and scatter outward as a result of elastic binary collisions with surface atoms. 

Inelastic energy losses are neglected in the original model by \citet{Forlano_1996}. We include inelastic energy losses (yellow overlay in Fig.\ref{fig:scattering_schematic}) by modifying the high-energy asymptotic limit of the model. In the discrete-streams formalism, the quantity $F_2\left(z=0, s\gg1\right)$ \citep[Eqs.~6 and 16]{Forlano_1996} is the Mellin transform \citep[for example,][]{Bertrand:1995aa} of the energy distribution at depth $z=0$ of particles backscattering with the highest energies ($s\gg1$). These particles follow the most efficient path consisting of only two elastic collisions with surface atoms (given an average scattering angle of $45^\circ$) to reverse from normal incidence, resulting in a maximum $\emph{elastic}$ energy of $p^2E_\mathrm{in}$, with $p$ the relative energy loss during an elastic binary-collision. We model inelastic losses by reducing this cut-off energy to $p^2E_\mathrm{in} - \Delta_\epsilon$, where $\Delta_\epsilon$ is the \emph{total} inelastic energy loss for particles on this direct trajectory. This leads to a modified asymptotic expression \citep[see Eq.~16]{Forlano_1996}:

\begin{equation}
    F_2\left(z=0, s\gg1\right) \propto \left(p^2E_\mathrm{in} - \Delta_\epsilon\right)^{s-1}\;,
    \label{eq:asymptotic_limit}
\end{equation}

where the proportionality constant, describing the nuclear scattering probability, remains unchanged. This modification only affects the high-energy part of the distribution, and preserves the analytical form of the original solution. More practically, and for the case of velocity-proportional stopping power $S_e \propto \sqrt{E}$ (see Eq.~\ref{eq:electronic_stopping_power}) that varies slowly at high energies, we can express $\Delta_\epsilon \approx n_\mathrm{eff} \, S_e\left(E_\mathrm{in}\right) L_{\mathrm{eff}}$, with $n_\mathrm{eff}$ the mean atomic number density of the regolith, and $L_{\mathrm{eff}}$ the path length of this two-collisions scattering trajectory. By including this additional loss factor, the energy distribution of a projectile $\mathrm{P}$ scattering from a single-species surface of atoms $\mathrm{S}$ can be expressed as

\begin{subequations}
    \begin{align}
    \mathcal{J}_E^{\mathrm{sc}}\left(E;E_\mathrm{in}, U,\Delta_\epsilon\right) &= \begin{cases}
          \dfrac{2E\kappa}{\mathtt{n}\left(E+U\right)^2} f\left(x\right),\quad &\text{if } x > 0 \\[.4cm]
          0, \quad &\text{otherwise},\\
    \end{cases}\\[0.5cm]
    f\left(x\right) &= \dfrac{\mathcal{I}_2\left(x\right)}{x e^x}\;,\\
    x&=\kappa\ln\zeta\;,\\
    \zeta&=\dfrac{p^2\left(E_{\mathrm{in}}+U\right) - \Delta_\epsilon}{E+U}\;,\\
    \kappa&= \dfrac{mp\left(1-p\right)^{m-1}}{\gamma^m}\;,
    \end{align}
    \label{eq:scattered_model}
\end{subequations}

where $\mathcal{I}_2$ is the modified Bessel function of the first kind; $\gamma$ is the maximum fraction of energy transferred from a projectile to a surface atom during a single elastic collision as defined in Eq.~\ref{eq:gamma}; $U$ is the surface binding energy; and $m$ a factor that controls the power-law scattering cross section as given by Eq.\ref{eq:m}. To improve readability, we omitted the subscript ${\left(\mathrm{P}\rightarrow\mathrm{S}\right)}$ for $p,m, \Delta_\epsilon$, and $\gamma$ in Eqs. \ref{eq:scattered_model}.d and \ref{eq:scattered_model}.e. The constant $p_{\left(\mathrm{P}\rightarrow\mathrm{S}\right)}$ defines the relative energy lost during a single elastic collision that deflected the projectile by $45^\circ$, and is defined as in \citet[Eq. 8]{Forlano_1996}:

\begin{equation}
    p_{\left(\mathrm{P}\rightarrow\mathrm{S}\right)} = \dfrac{M_{\mathrm{P}} \sqrt{2M_{\mathrm{S}}^2-M_{\mathrm{P}}^2} + M_{\mathrm{S}}^2}{\left(M_{\mathrm{P}} + M_{\mathrm{S}}\right)^2}\;.
\end{equation}

The constant $\mathtt{n}$ normalizes the function to unity, such that $\int_0^\infty \mathcal{J}_E^{\mathrm{sc}}\left(E\right)\mathrm{d}E = 1$. To reduce the computational cost of estimating the constant $\mathtt{n}$ via numerical integration, we evaluate it in an energy-normalized space, for which an approximate solution is provided in Appendix~\ref{appendix:scattered_norm_cst}. The function $f$ is computationally expensive, but can be well approximated for $10^{-5}< x < 10^2$ with a maximum relative error of 4\% by:

\begin{equation}
    f\left(x\right) \approx \dfrac{x/8}{\left(1 + 0.71521x + 0.32953x^2\right)^{1.28}}\;.
\end{equation}

\subsubsection{Elastic scattering}

\citet{Forlano_1996} approximates the interatomic potential by a power law controlled by an exponent $m=1/s$, with $s$ as defined in \citet[Eq. 3.5]{Lindhard1968APPROXIMATIONMI}. The special value $m=1$ describes the case of hard-sphere interaction, where electronic screening is negligible and the scattering approaches a pure Rutherford-type interaction, well-represented by the unscreened Coulomb potential \citep[Fig. 1]{Lindhard1968APPROXIMATIONMI}. In the case of a light projectile with an energy of hundreds of eV impacting on a heavier material, electronic screening cannot be neglected, that is $m<1$. We calculate the exponent $m_{\left(\mathrm{P}\rightarrow\mathrm{S}\right)}\left(E_\mathrm{in}\right)$ for a projectile $\left(M_\mathrm{P}, Z_\mathrm{P}\right)$ at energy $E_\mathrm{in}$ impacting a surface atom $\left(M_\mathrm{S}, Z_\mathrm{S}\right)$ as

\begin{equation}
    m = 0.5 \left[1-\dfrac{\mathrm{d}\ln s_\mathrm{r}\left(\epsilon\right)}{\mathrm{d}\ln\epsilon}\right]\;,
    \label{eq:m}
\end{equation}

where $s_\mathrm{r}\left(\epsilon\right)$ is the reduced nuclear stopping cross section \citep[Eq. 4.11]{Lindhard1968APPROXIMATIONMI}, and $\epsilon$ is the reduced energy (see Eq.~\ref{eq:reduced_energy}) evaluated at $E_\mathrm{in}$. We calculate $s_\mathrm{r}$ using the Krypton-Carbon potential, well approximated by \citet[Eq. 14]{Ziegler1985}. Deriving $s_\mathrm{r}$ from a Thomas-Fermi potential \citep[Fig. 2, Table 2b]{Lindhard1968APPROXIMATIONMI} gives very similar results. 

To approximate the exponent $m$ for a hydrogen atom interacting with the multiple atomic species in the lunar regolith, we calculate an effective value as

\begin{equation}
m_{\left(\mathrm{H}\rightarrow\mathrm{eff}\right)} = \sum_\mathrm{S} r_\mathrm{S} \, m_{\left(\mathrm{H}\rightarrow\mathrm{S}\right)} \;,
\end{equation}

where $r_\mathrm{S}$ is the atomic fraction of species $\mathrm{S}$ in the lunar regolith, as listed in Table~\ref{tab:elemental_composition}. For a projectile energy of 470\,eV (300\,km/s), we obtain $m_{\left(\mathrm{H}\rightarrow\mathrm{eff}\right)} \approx 0.57$, and for 1300\,eV (about 500\,km/s), $m_{\left(\mathrm{H}\rightarrow\mathrm{eff}\right)} \approx 0.66$. As the projectile energy increases, the interatomic potential exponent $m$ approaches unity, consistent with a transition toward hard-sphere–like nuclear interactions.

\subsubsection{Energy straggling}

When a beam of mono-energetic ions travels through a material, individual ions do not lose the same amount of energy $\Delta_\epsilon$ through inelastic collisions. Instead, the inelastic energy loss follows a statistical distribution with mean $\mu_\epsilon$ and a width characterized by the standard deviation $\sigma_\epsilon$. We refer to this width as "energy straggling". 

The energy straggling is caused by statistical fluctuations in the inelastic collision processes. \citet{Kaneko_1990} showed that the excitation of outer-shell electrons plays an important role in the energy loss of slow ions. Using a shell-wise local electron density model for hydrogen--lead interactions, they reported \citep[Table 1]{Kaneko_1990} that the 6s and 6p shells dominate inelastic losses at low velocities, with relative straggling $\sigma_\epsilon/\mu_\epsilon$ of 16\% and 11\%, respectively. Inner shells contribute much less to the total energy loss at low velocities because slow projectiles rarely penetrate deeply enough to interact with them. Nevertheless, when such interactions do occur, the associated energy straggling can be large. For example, the straggling associated with the 2s shell is large, $\sigma_\epsilon/\mu_\epsilon = 560\%$. The lunar regolith is primarily composed of oxygen atoms, whose electronic configuration $1\mathrm{s}^2 2\mathrm{s}^2 2\mathrm{p}^4$ involves only such inner shells, all of which are associated with large energy straggling.

To introduce the energy straggling in our model, we treat $\Delta_\epsilon$ as a random variable drawn from a normal distribution truncated to the physically-allowed interval $0 \leq \Delta_\epsilon < \Delta_{\epsilon\max}$, with $\Delta_{\epsilon\max} = p^2 (E_\mathrm{in} + U) - U$ derived from the condition $x > 0$ (see Eq.~\ref{eq:scattered_model}):

\begin{equation}
    \Delta_\epsilon \sim \mathtt{TruncNorm}\left(\mu=\mu_\epsilon, \sigma=\sigma_\epsilon, \mathrm{lower}=0,  \mathrm{upper}=\Delta_{\epsilon\max}\right)\;.
\end{equation}

The expected value of the energy distribution $\mathcal{J}_E^{\mathrm{sc}}\left(E; \Delta_\epsilon, \ldots\right)$ over the truncated Gaussian can be expressed as

\begin{subequations}
    \begin{align}
        \mathbb{E}\left[ \mathcal{J}_E^{\mathrm{sc}}\left(E; \Delta_\epsilon\right) \right] &= \frac{1}{Z\sqrt{\pi}} \int_{t_0}^{t_1} \mathcal{J}_E^{\mathrm{sc}}\left(E;\Delta_\epsilon= \mu_\epsilon + \sqrt{2} \sigma_\epsilon t\right) e^{-t^2} \mathrm{d}t\;, \\
        t_0 &= - \dfrac{\mu_\epsilon}{\sqrt{2}\, \sigma_\epsilon}\;,\\
        t_1 &= \dfrac{\Delta_{\epsilon\max} - \mu_\epsilon}{\sqrt{2}\, \sigma_\epsilon}\;,\\
        Z &= \mathtt{erf}\left(t_1\right) 
    - \mathtt{erf}\left(t_0\right)\;,
    \end{align}
    \label{eq:scattered_model_full}
\end{subequations}

where $\mathtt{erf}$ is the error function, and $t$ the integration variable. In practice, this integral is evaluated numerically using the Gauss-Legendre quadrature restricted to the interval $[t_0,t_1]$:
\begin{subequations}
\begin{align}
    \mathbb{E}\left[ \mathcal{J}_E^{\mathrm{sc}}\left(E; \Delta_\epsilon\right) \right] &\approx \frac{t_1-t_0}{Z\sqrt{\pi}} \sum_i^N 
w_i \mathcal{J}_E^{\mathrm{sc}}\left(E, \Delta_\epsilon=\mu_\epsilon + \sqrt{2} \sigma_\epsilon t_i\right) e^{-t_i^2}, \\
    t_i &= \dfrac{x_i+1}{2}\left(t_1-t_0\right)+t_0\,, 
\end{align}
\label{eq:scattered_model_straggling}
\end{subequations}

where $x_i$ and $w_i$ are the $i$-th Gauss-Legendre root and weight of the $N$-th Legendre polynomial, respectively. For a sufficiently large $N$, the property $\int_0^\infty \mathbb{E}\left[ \mathcal{J}_E^{\mathrm{sc}}\left(E; \Delta_\epsilon\right) \right] \mathrm{d}E = 1$ is conserved. To make the notation clear, we write the energy distribution of a projectile $\mathrm{P}$ that scatters from surface atoms $\mathrm{S}$ as

\begin{equation}
\mathcal{J}_{\left(\mathrm{P}\rightarrow\mathrm{S}\right)}^{\mathrm{sc}}\left(E;E_\mathrm{in}, U,\mu_\epsilon, \sigma_\epsilon\right) = \mathbb{E}\left[ \mathcal{J}_E^{\mathrm{sc}}\left(E;E_\mathrm{in}, U,\Delta_\epsilon\right) \right] .
\label{eq:scattered_model_final}
\end{equation}

\subsubsection{Multi-species surface}

To extend the model from \citet{Forlano_1996} to a multi-species surface, a rigorous approach would be to modify the energy transfer function $K$ in \citet[Eq. 1]{Forlano_1996}. However, to preserve the analytical traceability of the original solution, we decided to weight the contribution of each atomic species to the reflection process by directly weighting Eq.~\ref{eq:scattered_model_final}, that is

\begin{equation}
    \mathcal{J}_{\left(\mathrm{P}\rightarrow\mathrm{all}\right)}^{\mathrm{sc}} \equiv \sum_{\mathrm{S}} w_{\mathrm{S}} \cdot \mathcal{J}_{\left(\mathrm{P}\rightarrow\mathrm{S}\right)}^{\mathrm{sc}}\;,
\end{equation}

with $w_{\mathrm{S}} \equiv r_\mathrm{S} \; S_n^{\left(\mathrm{P}\rightarrow \mathrm{S}\right)}\left(E_\mathrm{in}\right)$, where $S_n$ is proportional (up to a species-independent factor) to the nuclear stopping power. This weighting is analogous to the species-dependent weighting factor used in \citet{Eckstein_2003} when modelling the energy dependence of sputtering yields. $S_n$ is defined as \citep[Eq. 9]{Wilson1977}

\begin{equation}
    S_n^{\left(\mathrm{P}\rightarrow \mathrm{S}\right)}\left(E\right) \propto \dfrac{M_\mathrm{P} Z_\mathrm{P} Z_\mathrm{S}}{\left(M_\mathrm{P} + M_\mathrm{S}\right) \sqrt{Z_\mathrm{P}^{2/3}+Z_\mathrm{S}^{2/3}}} s_\mathrm{r}\left[\epsilon_{\left(\mathrm{P}\rightarrow \mathrm{S}\right)}\left(E\right)\right]\;.
    \label{eq:total_stopping_power}
\end{equation} 

Weighting by the nuclear stopping power encodes the fraction of the total nuclear stopping power contributed by species $\mathrm{S}$, directly reflecting its ability to deflect an incoming particle $\mathrm{P}$.

Table \ref{tab:elemental_composition} lists, for a hydrogen atom as projectile, the nuclear stopping power, up to a constant, of the atomic elements that make up the lunar regolith. The values range from 5.08 to 6.88. Lighter elements show higher stopping powers, meaning they are more effective at deflecting hydrogen. Therefore, species heavier than oxygen--which dominates the atomic composition--play a minor role, both because of their lower abundance and their reduced ability to deflect hydrogen. Therefore, describing the surface as composed of a single effective species is reasonable

\begin{equation}
    \mathcal{J}_{\left(\mathrm{P}\rightarrow\mathrm{all}\right)}^{\mathrm{sc}} \approx \mathcal{J}_{\left(\mathrm{P}\rightarrow\mathrm{S_{eff}}\right)}^{\mathrm{sc}}\;,
    \label{eq:scattering_model_final2}
\end{equation}

with $\mathrm{S_{eff}} = \left(M_\mathrm{eff}, Z_\mathrm{eff}\right)$ defining the averaged atomic mass and atomic number of the lunar regolith, weighted by their atomic abundance. From Table~\ref{tab:elemental_composition}, we get $\mathrm{S_{eff}} = (21.88\;\mathrm{amu}, 10.81)$.

\subsubsection{Constraining the model with simulations}
\label{subsubsec:scattered_model_simulation_constraints}

The two main unknown parameters in our model are the mean inelastic energy loss, $\mu_\epsilon$, and the energy loss straggling, $\sigma_\epsilon$. The two quantities constrain the inelastic energy loss, $\Delta_\epsilon$, that is the \emph{total} energy lost inelastically by a particle as it travels through a material. A longer path length within the material leads to a greater total loss. We approximate the dependence of $\Delta_\epsilon$ on the path length of a particle in the material by expressing it as a function of the total scattering angle, $\Psi$. We define the total scattering angle as the angle between the incident particle direction and the emission direction (see Fig.~\ref{fig:all_angles}). Its relation to the incident and emission angles is 

\begin{equation}
\Psi= \pi - \arccos{\big[\cos{\left(\mathrm{SZA}\right)}\cos{\left(\beta\right)}+\sin{\left(\mathrm{SZA}\right)}\sin{\left(\beta\right)}\cos{\left(\phi\right)}\big]}\;,
\end{equation}

with $\Psi$, $\mathrm{SZA}$, $\phi$ and $\beta$ all in radians. For a particle that reverses its direction completely--backscattering along the incident path--the scattering angle is $\Psi = 180^\circ$. Scattering at a grazing incidence and emission corresponds to $\Psi \approx 0^\circ$. A larger scattering angle implies a longer path length within the material--a particle that reverses its direction of motion is likely to suffer many collisions, leading to a longer path length.

The total inelastic energy loss must also depend on the energy of the projectile, which increases with increasing incident energy. Therefore, we aim to parametrize $\mu_\epsilon\left(E_\mathrm{in}, \Psi\right)$ and $\sigma_\epsilon\left(E_\mathrm{in}, \Psi\right)$. To do so, we compare and constrain our models with SDTrimSP-3D simulations of solar wind protons scattering from a rough surface representing lunar regolith \citep{Szabo_2023}. These simulations aim at reproducing the energy dependence of the differential flux of emitted energetic neutral hydrogen atoms from the lunar surface and show good agreement with observations from the Chandrayaan-1 Energetic Neutral Analyzer (CENA) instrument \citep{barabash2009_sara_on_chandrayaan}. We first constrain the dependence on the total scattering angle using simulation results from \citet[Figs. 4 and 7]{Szabo_2023a}. We then constrain the dependence on the energy of the incident particle using \citet[Fig. 5b]{Szabo_2023a}.

\subsubsection{Angular-dependence of the inelastic loss and straggling}
\label{subsubsec:inelastic_loss_and_straggling}

We constrain the angular dependence using simulations from \citet[Figs.~4 and 7]{Szabo_2023a}, which show the energy distributions of protons scattered from a regolith-like surface as functions of the macroscopic emission angle $\beta$. The simulation data cover two different solar zenith angles (SZA): $0^\circ$ and $60^\circ$.

\begin{figure}[ht!]
  \resizebox{\hsize}{!}{\includegraphics{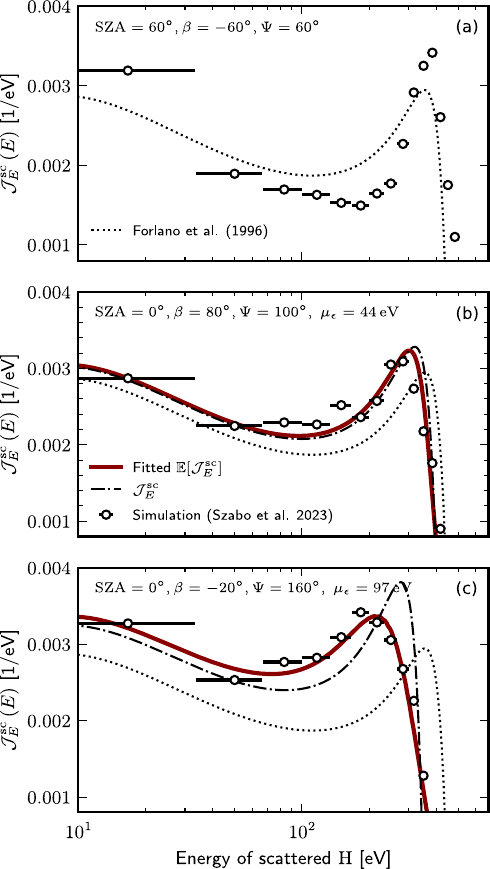}}
  \caption{Comparison of our energy distribution model with simulation data for 300\,km/s solar wind protons (470\,eV) scattering off lunar regolith at three different scattering angles. Panels a), b), and c) show the fitted model for three different total scattering angles $\Phi=60^{\circ}, 100^{\circ}$ and $160^{\circ}$, respectively. The original model from \citet{Forlano_1996} in panel (a) provides a reasonable match to the simulation data but generally overestimates the energy loss. The energy straggling is set to $\sigma_\epsilon = 0.7\,\mu_\epsilon$. For both panels (b) and (c): the original model (dotted line) from \citet{Forlano_1996} provides a reasonable average fit but consistently overestimates the energy of the high-energy scattered population. The modified model (Eq.~\ref{eq:scattered_model}; dash-dotted line), which includes a constant inelastic loss, improves the fit, though it underestimates the width of the high-energy peak at large scattering angles. Including energy loss straggling (Eq.~\ref{eq:scattered_model_final}, red solid line) further improves the fit. All comparisons use a surface binding energy of $U = 5$\,eV.}
    \label{fig:forlano_vs_sim_angles}
\end{figure}

We fit the parameters $(\mu_\epsilon, \sigma_\epsilon)$ in our model to match the simulated energy distributions for all available combinations of SZA and $\beta$, summarized by the total scattering angle $\Psi$. Figure ~\ref{fig:forlano_vs_sim_angles} compares our fitted model with the simulated spectra. We show, for comparison, models with (Eq.~\ref{eq:scattered_model_final}) and without (Eq.~\ref{eq:scattered_model}) energy straggling, and the original model of \citet{Forlano_1996}. 


Our model reproduces the simulated spectra well, except for total scattering angles smaller than about 100$^\circ$. In this regime illustrated in Fig.~\ref{fig:forlano_vs_sim_angles}a, the original model of \citet{Forlano_1996} already overestimates the energy loss, and adding an inelastic energy loss term further degrades the fit. A likely explanation for this poorer agreement at $\Psi < 100^\circ$ is that the model from \citet{Forlano_1996} was formulated for particles incident normal to the surface and scattering away from it, thereby restricting the scattering angle to $\Psi \in [90^\circ, 180^\circ]$. The model is not derived for smaller scattering angles.

\bigbreak

\begin{figure}[!ht]
  \resizebox{\hsize}{!}{\includegraphics{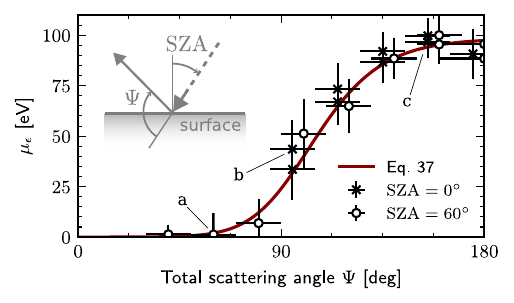}}
  \caption{Relation between the mean inelastic energy loss, $\mu_\epsilon$, and the total scattering angle, $\Psi$. The mean inelastic energy loss is fitted for different solar zenith angles (crosses: normal incidence; circles: 60$^\circ$ incidence) and for various emission angles. Crosses are shifted by -5$^\circ$ for clarity. Equation~\ref{eq:inelastic_loss_angular_dependence} is shown as the solid red line. The fits from the three panels in Fig.~\ref{fig:forlano_vs_sim_angles} are labelled. 
  }
\label{fig:inelastic_loss_angular_dependence}
\end{figure}

In a second step, we fit the parameters $(\mu_\epsilon, \sigma_\epsilon)$ obtained from fitting the simulated spectra in Fig.~\ref{fig:forlano_vs_sim_angles} as a function of $\Psi$. For $\mu_\epsilon$ we obtain the fit presented in Fig.~\ref{fig:inelastic_loss_angular_dependence}. From this study, we find that the mean inelastic energy loss, $\mu_\epsilon$ [eV], shows a sigmoid-curve dependence on the scattering angle, $\Psi$ [deg], given by:

\begin{equation}
    \mu_\epsilon\left(\Psi, E_\mathrm{in}=470\;\mathrm{eV}\right) \approx 94.6 \; \Big\{1+\exp\left[-0.0588 \left(\Psi - 90\right)\right]\Big\}^{-2.03}\,.
    \label{eq:inelastic_loss_angular_dependence}
\end{equation}

We observed that a constant relative energy loss straggling of $\sigma_\epsilon = 0.7 \mu_\epsilon$ fits the simulation data well.

\subsubsection{Energy-dependence of the inelastic energy loss and straggling}

\begin{figure}[ht!]
    \centering
    \includegraphics[width=1\linewidth]{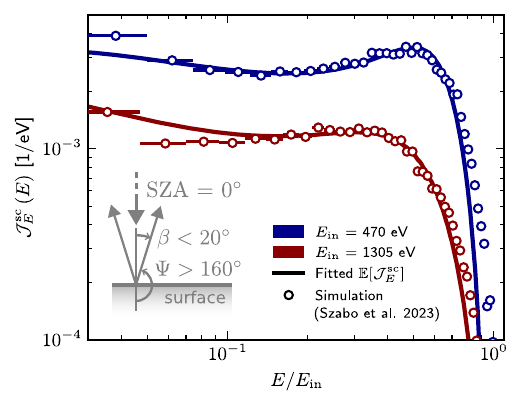}
    \caption{Comparison between the energy distribution model (Eq.~\ref{eq:scattered_model_final} solid lines) and simulated energy distributions (open circles) of solar wind proton scattering at normal incidence on the lunar regolith for proton incident speeds of 300\,km/s (blue) and 500\,km/s (red) \citep[Figs.~4 and 5]{Szabo_2023}. We assumed a surface binding energy of $U = 5$ eV.}
    \label{fig:forlano_vs_sim}
\end{figure}

The inelastic energy loss depends on the projectile energy: as the projectile energy increases, the particle loses more energy through inelastic processes (see Eq.~\ref{eq:electronic_stopping_power}) and generally travels longer within a material. In this section, we fit the mean inelastic energy loss, $\mu_\epsilon$, and the energy loss straggling, $\sigma_\epsilon$, of our model to the data presented in \citet[Figs. 4 and 5]{Szabo_2023}, corresponding to scattering angles $\Psi > 160^\circ$. These data are then used to constrain the amplitude of Eq.~\ref{eq:inelastic_loss_angular_dependence} as a function of the projectile energy. The fitted models, together with the simulated energy spectra, are shown in Fig.~\ref{fig:forlano_vs_sim}, and the resulting model parameters are summarized in Table~\ref{tab:scattering_model_parameters}. 

The model fits the simulated data reasonably well. The high-energy peak is accurately reproduced, while discrepancies appear at low energies corresponding to the multiple-scattering regime. One possible reason is that inelastic losses are included in our model only for particles scattering after a small number of collisions ($s \gg 1$ in Eq.~\ref{eq:asymptotic_limit}). As a result, inelastic losses in the multiple-scattering regime are likely underestimated.

\begin{table}[h!]
\caption{Parametrization of the energy distribution model for different incident particle energy. The relative $(1\sigma)$ error on the fitted $\mu_\epsilon$ is about 1\%.}
\label{tab:scattering_model_parameters}
\centering
\begin{tabular}{lcc}
\hline\hline
$E_\mathrm{in}$ [eV] & $\mu_\epsilon$ [eV] & $\sigma_\epsilon / \mu_\epsilon$ [\%] \\ \hline
470                            & 94.6                           & 65                                             \\
1305                           & 388                          & 65                                             \\ \hline
\end{tabular}
\end{table}

Based on the data of Table~\ref{tab:scattering_model_parameters}, we generalize the mean inelastic energy loss model of Eq.~\ref{eq:inelastic_loss_angular_dependence} to arbitrary incident energies by scaling the amplitude with energy. This yields the relation

\begin{equation}
    \mu_\epsilon\left(\Psi, E_\mathrm{in}\right) = \mathrm{A}_{470} \left(\dfrac{E_\mathrm{in}}{470}\right)^{k_0} \cdot \Big\{1+\exp\left[k_1 \left(\Psi - 90\right)\right]\Big\}^{k_2}\;,
    \label{eq:inelastic_loss_model}
\end{equation}

with fit parameters $\mathrm{A}_{470} = 94.6 \; [\mathrm{eV}]$, $k_0 = 1.44$, $k_1 = -0.0588\,[1/\mathrm{deg}]$, and $k_2 = -2.03$. The exponent $k_0$ is kept a fixed constant throughout this paper. At low energies, inelastic losses are typically proportional to the projectile velocity (Sect.~\ref{subsubsec:inelastic_losses_general}), corresponding to $k_0 = 0.5$.
In contrast, our fit yields a much steeper dependence ($k_0 \approx 1.5$), which can be understood as higher-energy projectiles travel longer distances within the material, leading to larger total inelastic losses. We found that a constant relative energy loss straggling of $65\%$ fits the simulated spectra well. We write 

\begin{equation}
    \sigma_\epsilon\left(\Psi, E_\mathrm{in}\right) = 0.7 \mu_\epsilon \left(\Psi, E_\mathrm{in}\right)\;.
    \label{eq:energy_straggling}
\end{equation}

\bigbreak

Simulations by \citet{Szabo_2023} show that electronic stopping represents the dominant energy loss mechanism for protons with typical solar wind energies scattering off lunar regolith, accounting for approximately 65\% of the total energy loss for 300 km/s protons. We find a similar result: from Equation~\ref{eq:inelastic_loss_model} and for $E_\mathrm{in} =470$\,eV (as an example), the maximum inelastic energy loss is about 90\,eV. In the single-collision regime (Eq.~\ref{eq:asymptotic_limit}), the minimum energy loss is given by $p^2 E_\mathrm{in} - \Delta_\epsilon$, where $\Delta_\epsilon \approx 90$\,eV corresponds to 71\% of the total energy loss for $E_\mathrm{in} = 470$ eV. For particles contributing to the high-energy peaks in Fig.~\ref{fig:forlano_vs_sim}, inelastic losses account for roughly 50\% of the total energy losses.

We can approximate the shortest path length of scattered protons using the following relationship: $L_{\mathrm{eff}} \approx \Delta_\epsilon / \left[ n_\mathrm{eff}\cdot S_e\left(E_\mathrm{in}\right)\right]$ (see Section \ref{adding_inelastic_losses}). Assuming an averaged electronic stopping power of $S_e \approx 0.2\;\mathrm{eV\,nm}^2$ (obtained from Eq.~\ref{eq:electronic_stopping_power} for an averaged regolith atomic number of 10.81 and for $E_\mathrm{in} = 470$\,eV), a regolith number density of approximately $n=28\;\mathrm{atoms}/\mathrm{nm}^3$ (assuming a true density of 3.035\,g/cm$^3$ and an averaged molar mass of 66\,g/mol), and a total inelastic loss of 90\,eV (for near $180^\circ$-backscattering protons), we obtain $L_\mathrm{eff} \approx 16$\,nm. \citet{Tucker_2019} showed that solar wind implants protons into the top 20-30\,nm of lunar regolith grains, a range consistent with our estimated path length. The estimated path length is also comparable to the hydrogen enhancement depth observed in regolith samples returned by Chang'e\,5 {\citep{Zhou2022ChangE-5-sample}}.



\section{Application to data}

The NILS data report the differential number flux of negative hydrogen ions emitted from the lunar surface as a result of precipitating solar wind ions. The observed flux, denoted $\mathcal{J}^-_\mathrm{H}$, depends on the properties of the precipitating flux, which varied moderately over the course of the NILS mission \citep{Wieser2025}. In the following, we briefly assess the impact of these variations and state the simplifying assumptions adopted in our model. 

During the mission, the solar zenith angle (SZA) varied from $52^\circ$ to $47^\circ$; a minimal variation over which we do not expect a significant change in $\mathcal{J}^-_\mathrm{H}$. For example, \citet{Broetzner2025} showed that the sputtering yield is largely independent of the angle of incidence. The angular distribution of the scattering and/or sputtering depends on the SZA \citep{Schaufelberger_2011}, although we neglect any dependencies on the SZA but in the estimation of $f_\perp$ and the inelastic energy loss model.

Similarly, the speed of the solar wind $v_{\mathrm{sw}}$ varied between $290\;\mathrm{km/s}$ and $310\;\mathrm{km/s}$. The yield $\eta^\mathrm{sc}$ depends on the incident energy of the solar wind protons and decreases with increasing energy \citep[Fig. 6b]{Lue_2018}, likely because deeper penetration reduces the probability of escape. Similarly, the sputtering yield depends on the precipitating energy. Nevertheless, given the restricted speed range covered during the NILS mission, we approximate $\eta^\mathrm{sc}\left(E_{\mathrm{sw}}\right)$ and $\eta^\mathrm{sp}\left(E_{\mathrm{sw}}\right)$ as constants $\overline{\eta^\mathrm{sc}}$ and $\overline{\eta^\mathrm{sp}}$.

The solar azimuth angle (SA, Fig. \ref{fig:all_angles}) angle varied from $43^\circ$E to $28^\circ$E, a total change of $15^\circ$. The NILS instrument provides an azimuthal coverage of approximately $15^\circ$ \citep[see $\Delta\alpha$ in Table 3]{CanuBlot2025}, which is comparable to this variation. Therefore, all quantities are given for an average azimuthal emission angle of $\overline{\phi} = 213^\circ$ (see Appendix~\ref{appendix:instrument_model} for further details). 

As indicated in Sect.~\ref{subsubsec:alpha_sputtering}, alpha particles are more efficient at sputtering hydrogen and solar wind protons. Although alpha particles do not significantly affect the energy distribution of sputtered hydrogen atoms, neglecting their contribution would underestimate the sputtered flux. We therefore account for their presence by scaling the precipitating proton flux with a correction factor $k_\alpha$, determined in in Sect.~\ref{subsubsec:alpha_sputtering}.

Starting from Eq.~\ref{eq:grand_model}, we therefore replace various quantities by their average values, explicitly state all dependences on the model parameters, and describe the observed negative ion flux as seen by NILS as follows

\begin{equation}
\begin{split}
    \mathcal{J}^-_\mathrm{H}&\left(E,\beta,\overline{\phi}\right) =\\
    &\mathrm{P}^-\left(v_\perp;v_\perp^{50\%},k\right) f_{\mathrm{sw}} \cos\left(\mathrm{SZA}\right)\;\cdot\\
    &\big\{\\
    &\quad\mathcal{J}_E^\mathrm{sp}\left(E;E_\mathrm{in}, U, \gamma_{\mathrm{extra}}\right)\; \left[\left(1+k_\alpha\right) \;\overline{\eta^\mathrm{sp}} \; \mathcal{J}_{\Omega}^\mathrm{sp}\left(\beta,\overline{\phi}\right)\right] \\
    &+\\
    &\quad\mathcal{J}_E^\mathrm{sc}\left(E;E_\mathrm{in}, U, \mu_\epsilon, \sigma_\epsilon \right) \; \left[\overline{\eta^\mathrm{sc}} \; \mathcal{J}_{\Omega}^\mathrm{sc}\left(\beta,\overline{\phi}\right)\right]\\
    &\big\}\;,\\
\end{split}
\label{eq:full_flux_model_applied_to_NILS}
\end{equation}

with $E_\mathrm{in} = E_\mathrm{sw}$;  $v_\perp = \cos\beta' \sqrt{2E/m_\mathrm{H}}$, where $m_\mathrm{H}$ is the proton mass and $\beta'$ the microscopic polar emission angle; $\mu_\epsilon = \mu_\epsilon\left(\Psi,E_\mathrm{in};\mathrm{A}_{470},k_1,k_2\right)$; $\sigma_\epsilon = 0.7\mu_\epsilon$; $\Psi=\Psi\left(\mathrm{SZA},\beta,\phi\right)$; and $k_\alpha = 0.11$.

\subsection{Additional prior knowledge needed} 
\label{sec:prior_knowledge}

The model in Eq.~\ref{eq:full_flux_model_applied_to_NILS} depends on the surface binding energy $U$ and the scattering/sputtering angular distributions, three quantities which are not well-constrained by the limited NILS data. Additionally, the model is close to non-identifiable \citep{Raue_2009} as both the scattering/sputtering yields and the probability of negative ionization influence the amplitude of the differential flux. A model is said to be non-identifiable when two or more parametrizations are observationally identical. We alleviate the non-identifiability issue and missing knowledge by statistically adding prior information, as described in the following sections.

We evaluated all priors using a prior predictive check \citep{Gelman_2014} by sampling from the joint prior $\mathbb{P}\left(\mathcal{J}_\mathrm{H}^-\right)$, propagating these samples through the model, and examining the resulting prior-predictive flux distribution. This ensured that the priors produced physically-plausible fluxes.

\subsubsection{Surface binding energy}

The surface binding energy (noted $U$ in the models) depends on the regolith base composition and absorbed species (hydroxylation of the regolith may affect the surface binding energy). We softly constrain the surface binding energy, shared by both the sputtering and scattering models, by using a truncated normal distribution ($\mathtt{TruncNorm}$) as a prior distribution:

\begin{equation}
    U\sim \mathtt{TruncNorm}\left(\mu=5,\; \sigma=0.5,\;\mathrm{lower=1}\right)\;,
\end{equation}

where all parameters are in units of eV. The probability is set to 0 for $U \le 1$. An average value of 5\,eV \citep{Kudriavtsev2005SurfaceBindungEnergy} was successfully used in earlier models \citep{Wieser_2024}. 

\subsubsection{Energetic neutral hydrogen albedo}

As stated in Sec. \ref{sec:prior_knowledge}, the model in Eq.~\ref{eq:full_flux_model_applied_to_NILS} has an identifiability issue, as both the scattering and sputtering yields and the probability of negative ionization influence the amplitude of the differential flux. The limited energy and angular coverage of the NILS data further complicates distinguishing the individual contributions.

To address this identifiability issue, we add the knowledge of a global lunar energetic neutral hydrogen albedo $\eta^\mathrm{ENA}$ of $0.16\pm0.05$, as derived in \citet{Vorburger_2013}. 
The study of \citet{Vorburger_2013} defines energetic neutral hydrogen atoms as hydrogen emitted from the lunar surface with energies above 11\,eV. We define this albedo using our model, given the following assumptions:

\begin{itemize}
\item[(1)] The albedo values reported by \citet{Vorburger_2013} are derived from the full observation of both angular distributions, $\mathcal{J}_\Omega^{\mathrm{sc}}$ and $\mathcal{J}_\Omega^{\mathrm{sp}}$.
\item[(2)] The neutralization probability $\mathrm{P}^0$ is assumed to be independent of both the emission energy and emission direction. Naturally, this cannot be as $\mathrm{P}^0 \approx 1 - \mathrm{P}^-$ and $\mathrm{P}^-$ depends on $v_\perp$. However, this approximation is reasonable as \citet{Wieser_2024} successfully predicted the energy distribution of the differential flux of neutral hydrogen atoms emitted from the lunar surface using the physical model of \citet{KENMOTSU_2004}, without considering any ionization probability. That model is closely related to the model of \citet{Ono_2005} adopted in the present study.
\item[(3)] The probability of positive ionization is small compared to the probability of neutralization. \citet[][Fig.~6.a]{Lue_2018} reported a charge ratio $\mathrm{H}^+/\mathrm{H}^0$ of approximately 5\% for hydrogen emitted at speeds similar as in our study. The ratio is derived from orbital observation, at an altitude at which we can assume all negative hydrogen ions have been converted to neutrals. We therefore write $\mathrm{P}^+ \ll \mathrm{P}^0 + \mathrm{P}^-$ and $\mathrm{P}^0 \approx 1 - \mathrm{P}^-$.
\end{itemize}

Based on the assumptions above, we obtain from Eq.~\ref{eq:grand_model}

\begin{equation}
    \eta^\mathrm{ENA} \equiv \mathrm{P}^0 \int_{11 \;\left[\mathrm{eV}\right]}^\infty \left[\eta^\mathrm{sc} \mathcal{J}_E^\mathrm{sc} + \left(1+k_\alpha\right) \;\eta^\mathrm{sp} \mathcal{J}_E^\mathrm{sp}\right]\mathrm{d}E\;,
    \label{eq:ENA_albedo}
\end{equation}

with $\mathrm{P}^0 = 1-\mathrm{P}^-$, where $\mathrm{P}^-$ is reduced to a constant equal to the average of $\mathrm{P}^-\left(v_\perp\right)$ for $v_\perp$ ranging from $10^4$ m/s to $10^5$ m/s. We added a constant $k_\alpha = 0.11$ to account for the alpha-induced sputtering, as in Eq.~\ref{eq:full_flux_model_applied_to_NILS}.

Practically, \citet{Vorburger_2013} were not able to reliably observe neutral hydrogen atoms with energies below 11\,eV. Using Eq.~\ref{eq:ENA_albedo}, we estimate the fraction of the total scattered and sputtered energetic neutral hydrogen observed by \citet{Vorburger_2013} to be 97\% and 59\%, respectively. Increasing the energy cut-off to 20\,eV reduces the observed sputtering fraction to 40\%; we nevertheless keep the  cut-off of 11\,eV to be able to compare results. Across the range of solar wind speeds considered (300–500\,km/s), the estimates change only by 3\%, a minimal variation that we neglect. The surface binding energy dependence has the largest effect, but it remains within $\pm10\%$, so we neglect it as well. This leads to the following prior parametrization of the yields

\begin{subequations}
    \begin{align}
        \eta^\mathrm{ENA} / \mathrm{P}^0 &\approx \eta^\mathrm{sc} + 0.6 \left(1+k_\alpha\right) \; \eta^\mathrm{sp}\;,\\
        \eta^\mathrm{sc} &\approx r \; \eta^\mathrm{ENA} / \mathrm{P}^0\;,\\
        \eta^\mathrm{sp} &\approx \left(1-r\right) \dfrac{\eta^\mathrm{ENA} / \mathrm{P}^0}{0.6 \left(1+k_\alpha\right)}\;,\\
        r &\sim \mathtt{Uniform}\left(0,1\right)\;,\\
        \eta_\mathrm{ENA} &\sim \mathtt{TruncNorm}\left(\mu=0.16, \sigma=0.05, \mathrm{lower}=0\right)\;,
    \end{align}
    \label{eq:ena_albedo}
\end{subequations}

where $\mathtt{Uniform}$ is a uniform distribution between the two given boundaries. 




\subsubsection{Angular prior distributions}

All quantities in the model of Eq.~\ref{eq:full_flux_model_applied_to_NILS} are either known or statistically constrained by prior information, with the exception of the angular emission profiles $\mathcal{J}_{\Omega}^\mathrm{sc}\left(\beta,\overline{\phi}\right)$ and $\mathcal{J}_{\Omega}^\mathrm{sp}\left(\beta,\overline{\phi}\right)$. 

The NILS instrument observed the lunar surface using an electrostatically-steerable angular pixel. The angular coverage of the surface is divided into eight intervals and covers emission polar angles from $\beta_{\max}=90^\circ$ (horizon-looking from the instrument perspective) to $\beta_{\min}=45^\circ$ (downward-looking from the instrument perspective). This angular scanning allows for extraction of information about the emission profiles from the NILS data. In the following, we statistically encode prior information of the scattered/sputtered emission profiles.

\paragraph{Scattering}

Prior knowledge of the angular distribution of scattered particles from the lunar surface is available from the simulation results of \citet[Fig.~1]{Szabo_2023}. Their simulated profile was obtained for solar zenith angles between $60^\circ$ and $75^\circ$, close to the average solar zenith angle of $50^\circ$ during the NILS mission. We parametrized this prior distribution as

\begin{subequations}
\begin{align}
&\mathcal{J}_\Omega^\text{sc, prior}\left(\beta\right) = \dfrac{f\left(\beta\right)}{2\pi}\;,\\[0.5cm]
&\begin{split}
    \text{with }f\left(\beta\right)
= &-0.22
+ 1.26\,\cos\beta
- 0.02\,\sin\beta\\
&- 0.29\,\cos\left(2\beta\right)
- 0.09\,\sin\left(2\beta\right)\;,
\end{split}
\end{align}
\label{eq:scattering_angular_prior}
\end{subequations}

with $\beta$ defined as in \citet{Szabo_2023}, that is, ranging from $-\pi/2$ to $\pi/2$, with forward-scattering for $\beta>0$ and backward-scattering (toward the Sun) for $\beta<0$. The profile is tilted such that about 40\% of emitted particles are forward-scattered. The NILS instrument observed the less-likely forward-scattering component of the emission profile, which is accounted for in the calculation of the total scattering yield.

\paragraph{Sputtering}

\citet{Cassidy_2005} simulated the sputtering of atoms from a porous regolith and showed that the angular emission of sputtered particles differs only little from that of a smooth surface. In both cases, the emission follows a cosine-law distribution. Alternatively, \citet{Jaeggi2024} showed that for oblique incidence of precipitating solar wind protons onto regolith-like surfaces, sputtering of heavy elements (for example, oxygen) is preferentially forward-directed, that is, along the solar wind flow. However, we expect hydrogen recoils to be produced nearly isotropically because the projectile and target masses are identical. Accordingly, we adopt a priori the following angular distribution:

\begin{equation}
\mathcal{J}_\Omega^\text{sp, prior}\left(\beta\right) = \dfrac{\cos\beta}{\pi}\;.
\label{eq:sputtering_angular_prior}
\end{equation}

\subsection{The near-surface lunar environment}
\label{sec:near_surface:environment}

Particle emission from lunar regolith is modulated by environmental parameters such as regolith composition at the landing site, local topography, surface disturbances caused by the lander, solar wind plasma conditions, and nearby magnetic anomalies. In the following section, we discuss how each of these factors affects the analysis of NILS data.

\subsubsection{Effect of the solar wind parameters}

We model the solar wind proton energy as Gaussian-distributed, with the mean equal to the bulk energy. The temperature of solar wind protons affects the precipitating energy distribution, which leads, in first-order, to a broadening of the scattered/sputtered energy distributions. 

Following \citet[Eq. 3]{Szabo_2023}, the standard deviation can be approximated as $\sigma_{\mathrm{sw}} \approx \sqrt{2E_{\mathrm{sw}}k_B T_{\mathrm{sw}}}$, with $k_B T_{\mathrm{sw}} \approx 1.5 \;\mathrm{eV}$ for $v_{\mathrm{sw}} = 300$ km/s, as expressed in \citet[Eq. 4]{Szabo_2023}.

We find that for solar wind speed between 300 km/s and 500 km/s, the temperature has only a minor effect on the energy distribution of scattered/sputtered particles. Such an effect is neglected, and we thereafter treat solar wind protons as mono-energetic. We additionally assume a uniform solar wind, such that all protons impinge on the surface at a constant SZA, approximated by the Sun direction, ignoring any aberration. We note that the Moon was outside of Earth's bow shock in undisturbed solar wind during the time NILS recorded data. We also neglect any heating from lunar magnetic anomalies as well as a possible deflection of the solar wind flow direction before it impinges onto the surface.

\subsubsection{Effect of magnetic anomalies}
\label{subsubsec:magnetic_anomalies}

The Chang’e-6 landing site is located west of a large cluster of lunar crustal magnetic anomalies near the South Pole–Aitken (SPA) basin. The direct interaction between the lunar crustal fields and the solar wind deflects and reflects incident solar wind protons \citep{futaana2003, Lue_2011}, reducing proton precipitation within magnetic anomalies and enhancing it in surrounding regions \citep{Wieser2010_ena_image, Vorburger_2013, Fatemi_2015, Maynadie2025}. The magnetic anomaly cluster near the SPA basin is the largest and among the strongest magnetic anomalies on the Moon. The high reflected proton densities of the cluster generate macroscopic plasma disturbances \citep{halekas2014, fatemi2014a, halekas2017a} disturbing proton precipitation over thousands of kilometres around the cluster \citep{Maynadie2025}. As a result, the precipitating solar wind proton energies and fluxes at the Chang'e-6 landing may significantly differ from upstream conditions.

To investigate these effects, we model the energy and flux of solar wind protons impacting the lunar surface observed by NILS as

\begin{subequations}
    \begin{align}
      f_\mathrm{sw} &= \nu_f \;f_\mathrm{sw}^\mathrm{up}\;,\\[0.2cm]
      E_\mathrm{sw} &= \nu_E\;E_\mathrm{sw}^\mathrm{up}\;,
    \end{align}
\end{subequations}

where $(f_\mathrm{sw}, E_\mathrm{sw})$ are defined as in Eq.~\ref{eq:full_flux_model_applied_to_NILS}, and $(f_\mathrm{sw}^\mathrm{up}, E_\mathrm{sw}^\mathrm{up})$ are the \emph{undisturbed} upstream solar wind flux and energy. Upstream solar wind parameters were obtained from measurements by the electrostatic analyzer \citep{McFadden:2008aa} onboard of the ARTEMIS-P2 spacecraft \citep{Angelopoulos:2011aa}. ARTEMIS-P2 was in an elliptic orbit around the moon at a distance of $10^3$--$10^4$\,km from NILS. The time-dependent multiplicative factors $(\nu_f, \nu_E)$ quantifying the effect of magnetic anomalies on the upstream solar wind flux and energy, are estimated in the next paragraphs. 

\paragraph{Simplified model}

The high-energy part of the scattered energy distribution mainly comprises of solar wind protons which only experienced few collisions with surface atoms. These reflected particles provide the most direct surface-originating proxy for the solar wind proton flux and energy impinging on the surface. 

We construct a simplified model for the high-energy ($>100\;\mathrm{eV}$) negative hydrogen differential flux $\mathcal{J}_\mathrm{H}^-$ emitted from the surface. We neglect the contribution from sputtering (minimal at these energies, as can be seen in Fig.~\ref{fig:NILS_surface_data}) and assume a constant ionization probability  $\mathrm{P}^-$ independent of emission energy and angles. The simplified model writes as

\begin{equation}
\begin{split}
        \mathcal{J}_\mathrm{H}^-\left(E\right) \propto \; &\nu_f  f_\mathrm{sw}^\mathrm{up} \cos\left(\mathrm{SZA}\right) \\[0.25cm]
        &\cdot \mathcal{J}_E^\mathrm{sc}\left[E; E_\mathrm{in} = \nu_\mathrm{E}  E_\mathrm{sw}^\mathrm{up}, U, \mu_\epsilon, \sigma_\epsilon\right] \\[0.25cm] 
        &\cdot \mathcal{J}^{\mathrm{sc,prior}}_\Omega\;,
\end{split}
\label{eq:SPA_effect}
\end{equation}

where the proportionality accounts for the scattering yield $\eta^\mathrm{sc}$ and the probability of ionization $\mathrm{P}^-$, both of which do not affect the inference of the variations of the factor $\nu_f$ and can be factored out for simplicity. For simplicity, we model the angular distribution of scattered hydrogen atoms using the prior distribution defined by Eq.~\ref{eq:scattering_angular_prior}. The inelastic energy loss $\mu_\epsilon$ is simplified to a constant inferred from data, with an energy straggling of $\sigma_\epsilon=0.7 \; \mu_\epsilon$ (Eq.~\ref{eq:energy_straggling}). The surface binding energy is set to $U=5\;\mathrm{eV}$ \citep{Kudriavtsev2005SurfaceBindungEnergy}. The factor $\nu_f$ is constrained by a weakly-informative prior (Table \ref{tab:priors}) that adds the knowledge that magnetic anomalies are unlikely to more than double the upstream solar wind flux \citep{Maynadie2025, Fatemi_2015}. 

While Eq.~\ref{eq:SPA_effect} constrains  the relative temporal variation of $\nu_f$, its absolute value remains unconstrained because the model is expressed up to a proportionality constant. However, \citet[Fig.~3]{Maynadie2025} showed that although magnetic anomalies significantly affect proton precipitation at the NILS location, the effect is highly variable and time-dependent, and the average precipitating proton flux remains comparable to that in regions not influenced by magnetic anomalies. Therefore, we assume that the average of $\nu_f$ computed  over the whole NILS mission--a duration of about one Earth day--provides a reasonable estimate for \emph{undisturbed} solar wind conditions, that is, $\overline{\nu_f} \approx 1$.

\paragraph{Inference from data}

\begin{figure*}
\sidecaption
  \includegraphics[width=12cm]{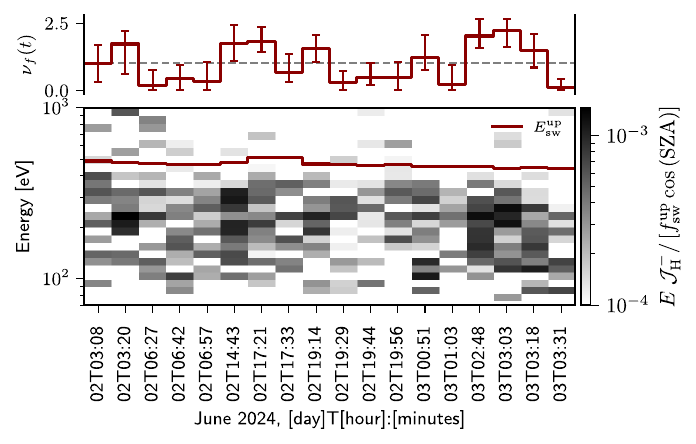}
    \caption{Energy-time spectrogram of the negative hydrogen ion energy-differential flux at the surface, normalized by the upstream solar wind flux. The flux is expressed in units of 1/(sr). The flux is averaged over angles. The time is discontinuous, an each bin is labelled by its average date (UTC). Lower panel: the red solid line represents the upstream energy. Upper panel: temporal variation of the solar wind flux enhancement/reduction factor, with the red horizontal line the best estimate (maximum-a-posteriori, \texttt{MAP}) and the vertical error bars the uncertainty (68\% highest density interval, \texttt{HDI}) (see Section \ref{sec:marginal_posterior}). Increases in the observed normalized flux are well reproduced by the model, corresponding to increases in the factor $v_f$.}
    \label{fig:energy_time_spectrogram}
\end{figure*}

Figure~\ref{fig:energy_time_spectrogram} shows the observed negative hydrogen ion energy-differential flux originating from the surface, noted as $E\; \mathcal{J}^-_\mathrm{H}$, normalized by the upstream flux $f_\mathrm{sw}^\mathrm{up}$. If the solar wind was not affected by the lunar environment, no variation in the observed flux would be expected beyond that arising from counting statistics. However, this is not what is observed: some time segments (for example, 02T19:44) show low fluxes, whereas others (for example, 03T03:03) show high fluxes. These variations are well-reproduced by the factor $\nu_f$, inferred from our simplified model. 

Figure~\ref{fig:energy_time_spectrogram} also shows the energy distribution of the energy-differential flux. The distribution shows temporal shifts in energy, with lower mean energies at, for example, 03T03:31 and higher mean energies at, for example, 03T02:48. Nevertheless, the energy distribution is rather constant and bounded by the energy of upstream solar wind protons, as represented by the red line in the lower panel of Fig.~\ref{fig:energy_time_spectrogram}. The factor $\nu_E$ could not be inferred due to non-identifiability versus $\nu_f$. For simplicity, we fix $\nu_E = 1$ hereafter. This approximation is justified by the limited magnitude of the observed energy shifts. The factor $\nu_E$ is likely less than one, that is, solar wind protons are systematically decelerated by magnetic anomalies. Such observation has already been reported by previous studies \citep{saito2012, futaana2013a, Fatemi_2015}.

We visually inspected the correlation between the scale factor $\nu_f$ and the following upstream conditions: solar wind proton density and speed; interplanetary magnetic field magnitude and direction. We did not observe any convincing correlation, supporting the idea of a turbulent propagation of the solar wind down to the surface. 

\subsubsection{Lunar regolith composition}

The lunar soil at the landing position of Chang'e-6 shows a bulk density of 0.983 g/cm$^3$ \citep{Li2024}, lower than most of Apollo, Luna, and Chang'e-5 samples, which typically ranges from 1.1 to 2.1 g/cm$^3$. This suggests the Chang'e-6 lunar soil is more porous than previously visited soils. The average density of individual grains is similar than the Chang'e-5 mission, at a value of 3.035 g/cm$^3$.

From the chemical composition of the \texttt{CE6C0000YJFM00102} and \texttt{CE6C0000YJFM00103} samples from Chang'e-6 \citep[Table 2]{Li2024}, we derive the average elemental composition of the lunar regolith at the Chang’e‑6 landing site. We calculated the relative atomic concentrations from the oxide weight percentages averaged over the two samples, assuming ideal stoichiometry and that all elements occur in the listed oxides. Beside the bulk composition, \citet{Lin_2025} also measured the hydrogen content in the outermost surface layer of the Chang’e-6 lunar grains. This region is particularly important for our study because the solar wind interacts mainly with the top few tens of nanometres of the regolith grains. Their measurements show that the soil grains are enriched in hydrogen of solar wind origin within the upper 200\,nm, with concentrations reaching up to 0.19 wt\% (equivalent of 1.7 wt\% $\mathrm{H}_2\mathrm{O}$ by weight), and increasing toward the surface. We add this extra solar wind-implanted hydrogen content to our regolith composition model, finally obtaining the first three columns of Table~\ref{tab:elemental_composition}. The elemental composition derived from the Chang’e-6 samples closely matches that reported by \citet[Table 2]{Wieser2025}.

\begin{table*}[h!]
\caption{Elemental composition of the regolith at the landing position of Chang’e-6, sorted by atomic mass. The model parameters are given for $\mathrm{H}$ interacting with each species $\mathrm{S}$.}
\label{tab:elemental_composition}
\centering
\renewcommand{\arraystretch}{1.1}
\setlength{\tabcolsep}{6pt}
\begin{tabular}{c c c c c c c c}
\hline\hline
\makecell{Element\\$\mathrm{S}$} &
\makecell{\tablefootmark{a}$(M, Z)$} &
\makecell{Abundance\\$r_\mathrm{S}$ [\%]} &
\makecell{$10^2\times C_{\left(\mathrm{H}\rightarrow\mathrm{S}\right)}$\\$[\mathrm{eV}\; \mathrm{nm}^2]$} &
\makecell{$10^3\times K_{\left(\mathrm{H}\rightarrow\mathrm{S}\right)}$\\ $[\mathrm{eV}^{1/2}\; \mathrm{nm}^2]$} &
\makecell{$\gamma_{\left(\mathrm{H}\rightarrow\mathrm{S}\right)}$} &
\makecell{$m$\\(470 / 1300 eV)} &
\makecell{$\propto S_n^{(\mathrm{H}\rightarrow \mathrm{S})}$\\(470 / 1300 eV)} \\
\hline
 H         & (1, 1)   &       4.1   &   4.136 &  4.299 & 1     & 0.76 / 0.80         & 5.85 / 3.21                \\
 O         & (16, 8)  &      58     &   5.232 &  8.701 & 0.22  & 0.59 / 0.69         & 6.88 / 5.20                \\
 Na        & (23, 11) &       0.18  &   5.502 &  9.225 & 0.16  & 0.55 / 0.65         & 6.48 / 5.29                \\
 Mg        & (24, 12) &       3.8   &   5.735 &  9.358 & 0.15  & 0.54 / 0.64         & 6.70 / 5.58                \\
 Al        & (27, 13) &       6.2   &   5.728 &  9.476 & 0.14  & 0.53 / 0.63         & 6.38 / 5.42                \\
 Si        & (28, 14) &      17     &   5.931 &  9.582 & 0.13  & 0.53 / 0.62         & 6.54 / 5.65                \\
 P         & (31, 15) &       0.022 &   5.922 &  9.678 & 0.12  & 0.52 / 0.61         & 6.25 / 5.48                \\
 K         & (39, 19) &       0.03  &   6.245 &  9.984 & 0.098 & 0.50 / 0.58         & 5.94 / 5.50                \\
 Ca        & (40, 20) &       4.7   &   6.395 & 10.05  & 0.095 & 0.49 / 0.58         & 6.00 / 5.62                \\
 Ti        & (48, 22) &       0.74  &   6.243 & 10.16  & 0.08  & 0.48 / 0.57         & 5.35 / 5.12                \\
 Mn        & (55, 25) &       0.068 &   6.381 & 10.3   & 0.07  & 0.47 / 0.55         & 5.08 / 4.99                \\
 Fe        & (56, 26) &       5.3   &   6.5   & 10.34  & 0.069 & 0.47 / 0.55         & 5.12 / 5.07                \\
\hline
weighted avg.        & (21.88, 10.81) &       100   &   5.484   & 8.902  &    0.21 & 0.57 / 0.66         & \textit{not relevant}                \\
 \hline
\end{tabular}
\tablefoot{Physical quantities for hydrogen interacting with each element S and in the last row the same quantities averaged using atomic abundances as weights. The elastic cross-section constant $C$ (Eq.~\ref{eq:elastic_cross_section}) and the inelastic energy loss constant $K$ (Eq.~\ref{eq:electronic_stopping_power}) are multiplied by a factor 100 and 1000, respectively, for clarity. The coefficient $m$ (Eq.\,\ref{eq:m}) and the total nuclear stopping power (up to a constant, Eq.\,\ref{eq:total_stopping_power}) are both given for 470 and 1300\,eV hydrogen interacting with each element. $\gamma$ is the maximum energy transfer factor during an elastic collision between a hydrogen projectile and a surface atom (Eq. \ref{eq:gamma}).\\
\tablefoottext{a}{$M$: mass in atomic mass unit, $Z$: atomic number.}
}
\end{table*}

\subsubsection{Lunar topography}
\label{subsec:lunar_topography}

\begin{figure}[ht!]
  \resizebox{\hsize}{!}{\includegraphics{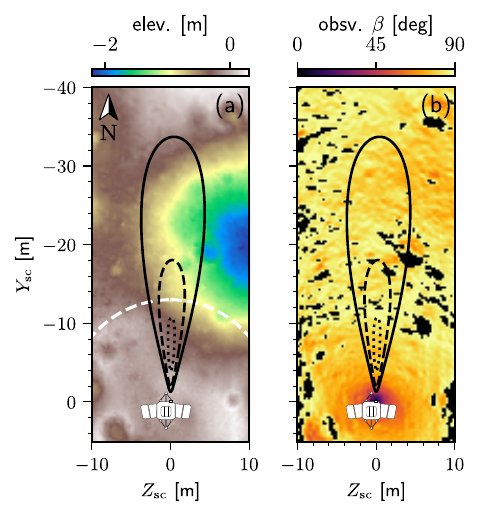}}
  \caption{Lunar topography and observed macroscopic polar emission angle near the Chang'e-6 landing site \citep{Liu2025}. Panel a: elevation relative to the lander. The estimated blast radius of the retro rockets of the lander is indicated by the white dashed line. Panel b: observed emission angle $\beta$. Regions that are occulted and not visible from NILS due to local surface topography are shown in black. In both panels, black contours indicate the sensitivity level of the instrument projected onto the surface:  20\% (solid line); 50\% (dashed line) and 90\% (dotted line). North is directed upward.}
  \label{fig:lunar_topology}
\end{figure}

The terrain surrounding the Chang’e-6 landing site was reconstructed from Landing Camera (LCAM) images with a spatial resolution of 1\,cm \citep{Liu2025}. The lander touched down on the rim of a 20\,m-wide crater. Using the resulting elevation model (Fig.~\ref{fig:lunar_topology}.a), we calculated the macroscopic emission angle $\beta$ as observed by the NILS instrument, which is mounted approximately 1.35\,m above the surface.
Figure~\ref{fig:lunar_topology}.b shows the emission angle $\beta$ across the lander coordinate system defined in \citet[Fig.~8]{CanuBlot2025}. The shown angle is an average over areas of 25\,cm$^2$. Despite the locally cratered terrain, the reconstructed polar emission angle differs only slightly from that of a flat surface. Across the NILS field of view, deviations from a flat surface remain within $\pm 5^\circ$, indicating that a flat-surface approximation is reasonable for our analysis.

\subsubsection{Regolith disturbances by the lander}

The exhaust plume of the lander engines mechanically disturbs the upper regolith layers at the landing site. For Chang’e-5, \citet{Zhang2022} reported an erosion depth of approximately 0.2 cm, corresponding to a total eroded mass of about 336 kg, with disturbances extending up to 13 m from the landing location. A large fraction of the area sampled by NILS lies within this radius (Fig.~\ref{fig:lunar_topology}a). Disturbed regolith is expected to show a modified grain-size distribution and a different solar wind exposure age \citep{1975Bibring}. Therefore, lander-based instruments observe a surface whose properties may differ from those inferred from orbital measurements, which average over larger and presumably pristine regions. The impact of these differences on the emitted negative ion flux, however, remains unknown.

\section{Inference from NILS measurements}
\subsection{Sampling of the posterior distribution}
\label{sec:posterior_sampling}

In this section, we apply the model in Eq.~\ref{eq:full_flux_model_applied_to_NILS} to the data from the NILS instrument to infer the following parameters: $\overline{\eta^\mathrm{sc}}$, $\overline{\eta^\mathrm{sp}}$, $\mathrm{A}_{470}$, $k_1$, $k_2$, $\mathrm{S}$, $v_\perp^{50\%}$, $k$, $U$, $\gamma_\mathrm{extra}$, $\mathcal{J}_{\Omega}^\mathrm{sc}\left(\beta,\overline{\phi}\right)$, and $\mathcal{J}_{\Omega}^\mathrm{sp}\left(\beta,\overline{\phi}\right)$.

To do so, we first constrain the model using prior knowledge of the relevant physical processes (see Sect.~\ref{sec:prior_knowledge} and Table~\ref{tab:priors}), and then update these priors with information from the NILS measurements through Bayesian inference \citep{ESTLER1999611,Gelman_2014}. The modelled differential flux $\mathcal{J}_\mathrm{H}^-$ is converted into an expected count rate using the instrument response described in Eq.~\ref{eq:instrument_model}. The prior knowledge $\mathbb{P}\left(\mathcal{J}_\mathrm{H}^-\right)$ is then updated using Bayes’ theorem (Eq.~\ref{eq:posterior_dist}), with a likelihood function that describes the (mass-separated) negative hydrogen ion count rate observed by the instrument. From the observed count rate $\vec{\mathcal{C}}$, we thus obtain the posterior distribution $\mathbb{P}\left(\mathcal{J}_\mathrm{H}^- \mid \vec{\mathcal{C}}\right)$, expressing our updated belief about the differential flux of negative hydrogen ions--and of all inferred parameters--given both the NILS data and our prior knowledge.

We follow the guideline proposed by \citet{Kruschke_2021} when describing the Bayesian inference. All posterior distributions were obtained through Markov chain Monte Carlo sampling \citep{PYMC,phan2019composable, bingham2019pyro,Hoffman_2011}.
The flux $\mathcal{J}_\mathrm{H}^-$ is described through our model in Eq.~\ref{eq:full_flux_model_applied_to_NILS} by a number of parameters (Table \ref{tab:post_summary}) aggregated in the vector $\vec{\theta} =\left[\overline{\eta^\mathrm{sp}}, \overline{\eta^\mathrm{sc}}, \mathrm{A}_{470}, \ldots\right]$. A set of $S=8\,000$ samples, $\{\vec{\theta}^{(s)}\}_{s=1}^S$, was drawn from the posterior distribution $\mathbb{P}\left(\vec{\theta} \mid \vec{\mathcal{C}}\right)$ using the \texttt{PyMC} Python package \citep{PYMC}. Four independent chains were run for 3\,000 iterations each, with the first 1\,000 iterations discarded as warm-up. We assessed convergence using the R-hat statistic \citep{Gelman_1992} and by visually inspecting the traces. We validated the model and the posterior by comparing the observed data to replicated data generated from the posterior predictive distribution.

Samples of the posterior distribution are shown in Fig.~\ref{fig:pair_plot}, where cross-correlations between parameters is visible. 

\subsection{Summary of the posterior distribution}

\subsubsection{Marginal posterior}
\label{sec:marginal_posterior}
\begin{table}[h!]
\caption{Summary of marginal posterior distributions for all model parameters.}
\label{tab:post_summary}
\centering
\renewcommand{\arraystretch}{1.3}
\begin{tabular}{ccrrrr}
\hline\hline
Parameter & Unit & Median & \texttt{MAP} & \multicolumn{2}{c}{68\% \texttt{HDI}} \\
 &  &  & & low & high \\
\hline
                $\overline{\eta^\mathrm{sc}}$                & &   0.215   &   0.222    &   0.161    &   0.271    \\
                $\overline{\eta^\mathrm{sp}}$                & &   0.106   &   0.0813   &   0.0378   &   0.156    \\
                     $\eta_\mathrm{ENA}$                     & &   0.240    &   0.243    &   0.197    &   0.278    \\
 $\overline{\eta^\mathrm{sc}} / \overline{\eta^\mathrm{sp}}$ & &   2.02    &   1.46     &   0.369    &   3.00        \\
                          $A_{470}$                          & eV & 147       & 147        & 131        & 163        \\
                            $k_1$                            & 1/deg &   0.0401  &   0.0382   &   0.0306   &   0.0471   \\
                            $k_2$                            & &   0.770    &   0.742    &   0.634    &   0.894    \\
                        $\mathrm{S}$                         & &   0.862   &   0.904    &   0.803    &   0.975    \\
                           $\left[v_\perp^{50\%} \times 10^{-6}\right]$\tablefootmark{a} & m/s & 12.0 &   9.41 & 5.35 & 16.7 \\
                             $k$                             & &   0.103   &   0.0998   &   0.0830    &   0.119    \\
                        $\mathrm{U}$                         & eV &   5.44    &   5.45     &   4.57     &   6.49     \\
                   $\gamma_\mathrm{extra}$                   & &   0.0613  &   0.0228   &   $\approx0$ &   0.0973   \\
\hline
\end{tabular}
\tablefoot{\tablefoottext{a}{For compactness, all values for $v_\perp^{50\%}$ have been scaled by $10^{-6}$. The original values can be recovered by multiplying the tabulated values by $10^{6}$.}}
\end{table}

Table~\ref{tab:post_summary} summarizes the \emph{marginal} posterior distributions of the model parameters estimated from the samples $\vec{\theta}^{(s)}$. Marginal quantities are obtained by integrating the posterior over all model parameters but the one under study. Such marginal distributions are also shown as red histograms in the diagonal panels of Fig.~\ref{fig:pair_plot}. The maximum-a-posteriori (\texttt{MAP})--the value of highest probability given the data and prior assumptions--of a parameter $\theta$ is estimated as

\begin{equation*}
    \theta_{\mathrm{MAP}} = \arg\max_{\theta}\,\mathbb{P}(\theta\mid\vec{\mathcal{C}})\;.
\end{equation*}

Uncertainty is quantified using the 68\% highest density interval (\texttt{HDI}), defined as the shortest interval $[a,b]$ satisfying
\begin{equation*}
    \int_a^b \mathbb{P}(\theta\mid\vec{\mathcal{C}})\,\mathrm{d}\theta = 0.68\;.
\end{equation*}
The \texttt{HDI} is computed from the marginal posterior samples and provides a robust uncertainty summary, particularly for skewed distributions.

\subsubsection{Joint-posterior}

The most complete representation of the posterior distribution is provided in Fig.~\ref{fig:pair_plot}, which displays the pairwise correlations between the parameters $\vec{\theta}$. The most relevant correlations are discussed below.

Figure~\ref{fig:pair_plot}.a shows a negative correlation between the scattering and sputtering yields. This is expected, as the energy distributions of scattered and sputtered particles overlap at low energies, and both yields are coupled through the scaling parameter $r$ (Eq.~\ref{eq:ena_albedo}).

Figures~\ref{fig:pair_plot}.b–c show a positive correlation between the individual yields and the ENA albedo, $\eta_\mathrm{ENA}$, consistent with the fact that the albedo is defined as the sum of the two yields.

Figure~\ref{fig:pair_plot}.at shows a weak positive correlation between $\gamma_\mathrm{extra}$ and the scattering yield. This is expected because $\gamma_\mathrm{extra}$ controls the high-energy cut-off of the sputtered energy distribution: a lower cut-off requires a higher scattering yield to reproduce the observed high-energy flux. For similar reasons, Fig.~\ref{fig:pair_plot}.ay shows an anti-correlation between $\gamma_\mathrm{extra}$ and $k_2$, which governs the width of the high-energy scattered peak.

All other panels are well behaved and show no strong correlations, indicating that the model parameters are identifiable.

\section{Data interpretation and discussion}

Figure \ref{fig:NILS_surface_data} shows the energy spectrum of the differential number flux of negative hydrogen ions (vertical bars) as measured by NILS for different observation geometries, along with the posterior distribution of the flux (lines). The differential flux (vertical bars) is derived only from measurements and knowledge of the instrument response and is independent of any physical model. The two agree well. High-energies are dominated by scattered particles, low-energies are populated with both sputtered and scattered particles. The domination of the scattered particles decreases as the emission polar angle increases. 

We note that the high-energy cut-off of the scattered energy distribution appears to be overestimated, which may indicate that the assumption $\nu_E = 1$ (Sect.~\ref{subsubsec:magnetic_anomalies}) is not representative. Constraining $\nu_E$ better will likely improve the agreement.

\subsection{An efficient ionizing material}

\begin{figure}[ht!]
    \centering
    \includegraphics[width=1.\linewidth]{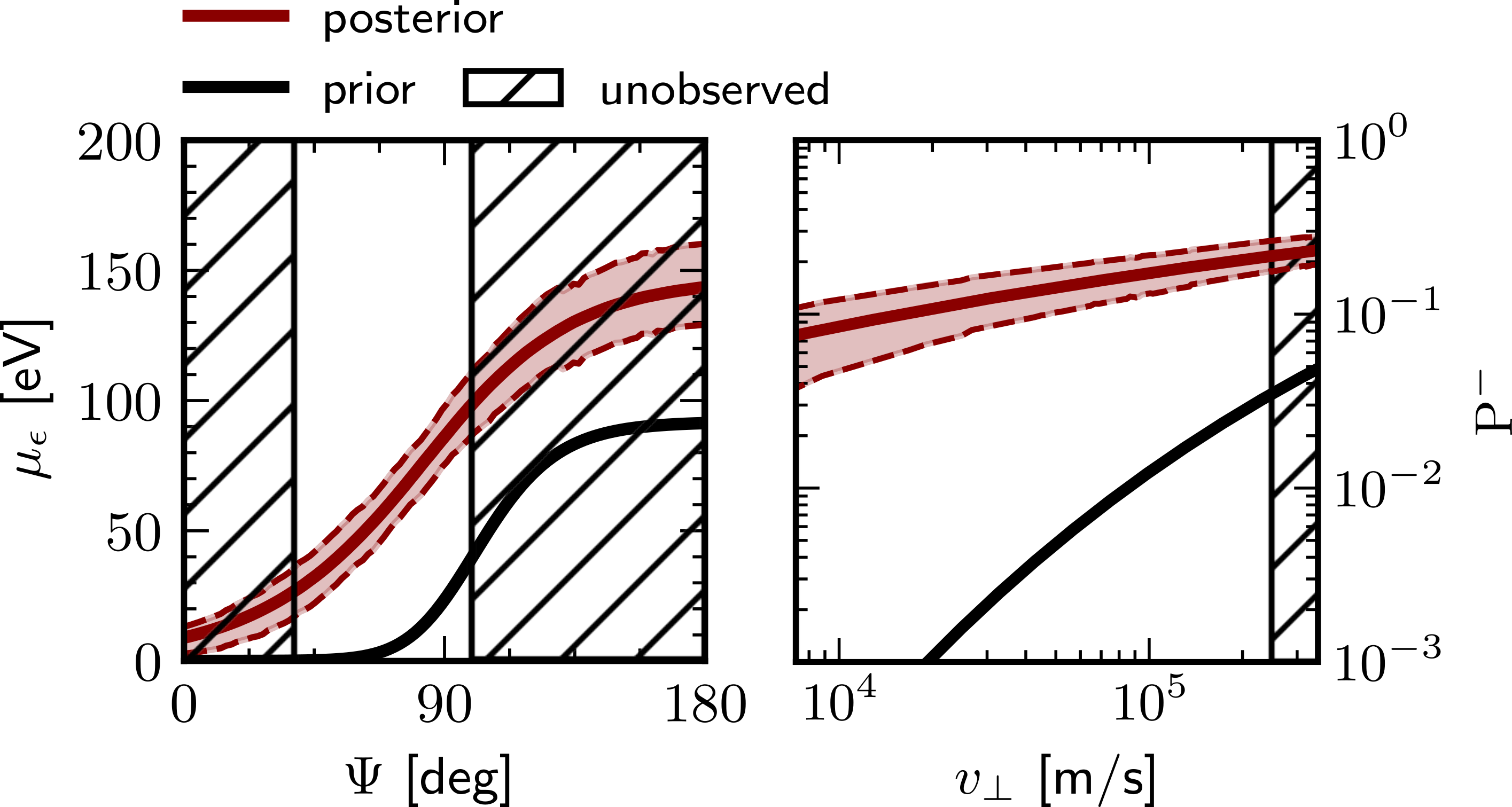}
    \caption{Comparison between the priors of the inelastic mean energy loss $\mu_\epsilon$ and the negative ionization probability $\mathrm{P}^-$, and their posterior distributions given the NILS data.}
    \label{fig:some_more_posteriors}
\end{figure}

The data shows that the lunar regolith is an efficient ionizing material for negative ions, with large probabilities of negative ionization, $\mathrm{P}^- > 10\%$ (Fig.~\ref{fig:some_more_posteriors}). The inferred probability shows a weaker dependence on the perpendicular velocity $v_\perp$ than suggested by our prior knowledge.

We tested the sensitivity of this result to the priors on $\mathrm{S}$, $v_\perp^{50\%}$, and $k$, and found the conclusion to be robust: the probability of negative ionization remains high and only weakly dependent on $v_\perp$. Among these, the prior on $k$ has the strongest influence on the inference, without significantly changing our conclusion.

Although lunar regolith appears to be an efficient source of negative ions, no orbital observations of negative ions originating from the lunar surface have been reported. This absence is likely due to the short lifetime of the negative charge state. \citet{Wieser2025} showed that the density of negative hydrogen ions decreases exponentially with altitude, with a scale height of about 10\,km.

\subsection{Longer transport within the regolith grains}

The mean inelastic energy loss $\mu_\epsilon$ is larger than what is predicted by \citep{Szabo_2023}, as can be seen in Fig.~\ref{fig:some_more_posteriors}. The amplitude $A_{470}$ increased from 94.6\,eV (as given by its prior in Table~\ref{tab:scattering_model_parameters}) to 147\,eV (as given by its posterior in Table~\ref{tab:post_summary}). We propose two explanations for this increase. 

\begin{enumerate}
    \item[(i)] The factor $\nu_E$ is likely smaller than unity--that is, solar wind protons are decelerated by magnetic anomalies--which would improve the agreement between the prior and posterior of $\mu_\epsilon$. Indeed, $\nu_E$ and $\mu_\epsilon$ are correlated: allowing the data to constrain $\nu_E$ yields $\nu_E \approx 0.85$ and $A_{470} \approx 110,\mathrm{eV}$. Additionally, $\nu_E$ may be time-dependent; if this is not accounted for, the mean inelastic energy loss could be overestimated to compensate for the increased spread in the precipitating proton energies.
    \item[(ii)] Inelastic losses were underestimated in the simulations by \citet{Szabo_2023}. This likely reflects an underestimate of the total path-length of the simulated particles in the regolith grains, rather than an underestimation of the inelastic energy loss, as the latter is well understood for hydrogen interacting with the atomic constituents of lunar regolith.
\end{enumerate}

We believe that both arguments are at play.

\subsection{Updated sputtering and scattering yields}

Our model defines a yield as the number of emitted hydrogen atoms per precipitating solar wind protons. The charge state is determined by the probability of ionization. Regardless of the charge state of the emitted hydrogen, we find a scattering yield of 22\% and a sputtering yield of 8.1\%, giving a ratio of scattered to sputtered hydrogen of $1.5^{+1.5}_{-1.1}$. This indicates that a proton is more likely to scatter off lunar regolith than to sputter surface hydrogen atoms.

Notably, the ENA albedo defined in Eq.~\ref{eq:ENA_albedo} rises from 16\% to 24\%, a relatively large value compared to most previous observations \citep{Vorburger_2013}. \citet[Fig.~2]{Futaana2012} reported similar values, up to 30\%, though their measurements were limited to the lunar equator. This increase in $\eta_\mathrm{ENA}$ can be attributed to the combination of a relatively low prior probability of negative ionization and the high fluxes measured by NILS. We note that fixing $\eta_\mathrm{ENA} = 0.16$ and sampling from the model does not change our conclusion that the probability of negative ionization remains high.

The scattering yield for negative hydrogen ions can be estimated in the first order as $\mathrm{P}^-\eta^\mathrm{sc}\approx 0.15 \times 0.22 = 3.3\%$, matching the yield of $2.5\%$ determined by a different method reported in \citet{Wieser2025}. The higher yield found here likely results from the energy distribution model of \citet{Wieser2025}, which underestimates the scattered low-energy population.
The sputtering yield for negative hydrogen ions is estimated as $\mathrm{P}^-\eta^\mathrm{sp}\approx 0.1 \times 0.08 = 0.8\%$, with the lower ionization probability accounting for the lower average energy of the sputtered particle population.

Similarly, the scattering yield for protons can be estimated from \citet{Lue_2018}, who reported a positive ionization probability of about 5\% for scattered protons of hundreds of eV. Assuming that 5\% of scattered hydrogen is emitted as protons, we obtain a proton scattering yield of about 1.1\%, in agreement with the values reported by \citet{Saito_2008, Lue_2014, Lue_2018}.

\subsection{Angular distributions}

The NILS observations cover only a limited range of emission angles. Nevertheless, for an average azimuthal emission angle of $\overline{\phi} = 213^\circ$, the distribution of the polar emission angle $\beta$ can be retrieved for both scattered and sputtered hydrogen atoms. Note that these distributions refer to emitted hydrogen atoms regardless of their charge state.

\begin{figure*}
\sidecaption
  \includegraphics[width=12cm]{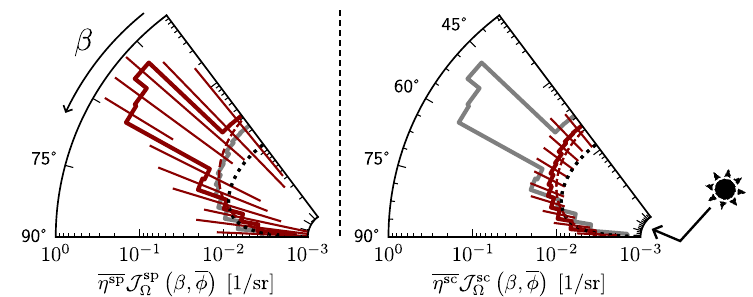}
    \caption{Posterior summary of the angular yields for sputtering (left) and scattering (right) as a function of the macroscopic polar emission angle $\beta$, at mean emission azimuth $\overline{\phi} = 213^\circ$. In both panels, the thick red line shows the \texttt{MAP} estimate of the (log) angular yield in each $\beta$ interval, with red radial lines indicating the 68\% \texttt{HDI}. For comparison, the grey lines show the complementary process (scattering in the sputtering panel; sputtering in the scattering panel). Dashed red curves indicate the priors (Eqs.~\ref{eq:scattering_angular_prior} and \ref{eq:sputtering_angular_prior}). The dotted black curve shows the (scaled) visibility function for lunar regolith (Fig.~\ref{fig:polar_emission_angle}). Yields correspond to forward-scattering and forward-sputtering, as reminded by the Sun symbol. Numerical values are listed in Table~\ref{tab:weights}.}
    \label{fig:NILS_angular_data}
\end{figure*}

Figure~\ref{fig:NILS_angular_data} shows the polar emission angle distributions for forward-scattered and forward-sputtered hydrogen. The angular distribution of scattered particles is better constrained by data, as indicated by the smaller uncertainties. In contrast, the sputtering distribution is less certain, primarily due to the limited instrument coverage at low energies. For both processes, the angular emission function for $\beta>75^\circ$ is controlled by visibility constraints. Regardless of the emission profile at the microscopic scale, the roughness of the regolith enforces a visibility reduction for near-grazing emission angles, as expressed by the dotted line in Fig.~\ref{fig:NILS_angular_data}. At smaller $\beta$ angles, the sputtered and scattered angular distributions begin to diverge.

The angular distribution of scattered particles closely matches its prior and shows little sensitivity to it, indicating that the prior provides an accurate physical description and agrees well with the data.

The angular distribution of sputtered particles rises sharply, with yields reaching up to 0.4\,sr$^{-1}$. The sputtering angular yield is sensitive to the width of the hierarchical scale prior, $\sigma_\mathrm{sp}$, defined in Table~\ref{tab:priors}. Narrowing this prior reduces bin-to-bin variations, effectively forcing the updated sputtering angular yield toward its a priori cosine dependence (Eq.~\ref{eq:sputtering_angular_prior}). However, this worsens the agreement between the modelled and observed differential flux, particularly for emission energies between 80\,eV and 200\,eV.

For large polar emission angles $\beta>75^\circ$, our observed angular emission function differs qualitatively from the one reported in \citet{Vorburger_2014} for energetic neutral hydrogen: while the former shows decreasing fluxes for large $\beta$, the latter does not show this. A possible cause for the difference is the lack of data coverage in \citet{Vorburger_2014} for $\beta$ close to 90$^\circ$, resulting in a weakly-constrained fit. 

It is important to note that, although the sputtering angular yields exceed the scattering angular yields at forward emission angles, the total scattering yield reported in Table~\ref{tab:post_summary} is higher than the total sputtering yield. This apparent discrepancy arises from the asymmetric scattering angular distribution, in which backward-scattering angles that are unobserved by NILS dominate and are constrained by the prior in Eq.~\ref{eq:scattering_angular_prior}. In contrast, the sputtering angular yield is assumed a priori to be symmetric and cannot be updated a posteriori given the limited angular coverage of NILS.

\subsection{Surface binding energy and other model parameters}

The surface binding energy $\mathrm{U}$ of regolith is another free parameter in our model. It appears not to be sensitive to the prior and its posterior robustly converges to about 4--6\,eV, with 5.5\,eV our best estimate. This is in good agreement with predictions \citep{Kudriavtsev2005SurfaceBindungEnergy}.

The extra energy loss factor $\gamma_\mathrm{extra}$ in the sputtering model shrunk towards zero in the posterior, indicating that neglecting this factor is compatible with the NILS data.

\subsection{Effects of lunar magnetic anomalies on the precipitating flux}

The solar wind flux precipitating onto the surface is not necessarily equal to the flux measured in nearby orbit. Variations in the precipitating flux of up to a factor of 1.5 have been observed on timescales of approximately 100\,s (Fig.~\ref{fig:energy_time_spectrogram}). The primary cause of these variations is the proximity to magnetic anomalies, which can partially shield \citep{Vorburger_2013} or locally enhance \citep{Wieser2010_ena_image} the precipitating flux. With a single-point measurement such as NILS, temporal and spatial variations cannot be disentangled; however, it is likely that both contribute. 

\subsection{Limitations and caveats}

Our model uses fixed values for the geometric factor of NILS without uncertainties for stability reasons. Any uncertainties in the geometric factor will directly propagate to additional uncertainties in the estimated probability of negative ionization for hydrogen $\mathrm{P}^-$. We obtain a relatively high values for $\mathrm{P}^-$ in our analysis. This could alternatively be caused by a systematic underestimation of the absolute geometric factor of NILS. A careful analysis shows that an underestimation of the geometric factor by a factor of two could be possible from an instrumental point of view, but is not very likely.

A NILS internal voltage offset on the electrodes controlling the angular response of the instrument limits the angular coverage at low energies. This offset is accounted for in this study to the best of our knowledge.

The NILS data are dominated by electrons despite the electron suppression system present in the instrument. This makes the extraction of the minor negative hydrogen ion signal statistically uncertain. The differential flux of electrons increases by orders of magnitude at low energies \citep[][Fig.~6.10]{CanuBlot_2024c}, further complicating the separation of the negative hydrogen ions. Although these effects are considered in this study, small inaccuracies in the calibration of the time-of-flight (or mass) responses for electrons and hydrogen ions could propagate into large uncertainties in the mass-separation, particularly at low energies. This may explain the relatively poor agreement between the model and the observed negative hydrogen ion differential flux below 50–60\,eV, as shown in Fig.~\ref{fig:NILS_surface_data}.

\section{Conclusion}

We developed and validated a semi-analytical model that describes the energy and angular distributions of negative hydrogen ions scattered and sputtered from the lunar surface. The model combines a physical description of the particle transport at the surface with a macroscopic description of the near-surface environment. A key feature of the model is that it treats scattered and sputtered fluxes separately and accounts for the probability of ionization of the emitted particles. 

The model is constrained by prior information on the relevant physical processes, informed by results from other instruments, simulations, and theoretical studies. We update these prior assumptions through Bayesian inference using the measurements obtained by the NILS instrument. 

The model agrees well with the NILS measurements. We observe that lunar regolith is an excellent ionizing material for negative ions, with a probability of a hydrogen atom to leave the surface as a negative ion of 7--20\%. We estimate that a precipitating solar wind proton has a 22\% chance of scattering and 8\% chance of sputtering a surface hydrogen, indicating a precipitating proton is more likely to scatter off lunar regolith than to sputter surface hydrogen atoms. 

We estimate that about 3.3\% of the solar wind protons are scattered as negative hydrogen ions, while about 0.8\% result in sputtering of a surface hydrogen atom as a negative ion.

We obtain a robust estimate of the surface binding energy $U$ of regolith of about 5.5\,V. We obtain a large inelastic energy loss, which may reflect that the total path-length of the particles in the regolith is larger than previously assumed. The surface roughness is found to control the angular emission distribution at near-specular angles for both the scattered and sputtered processes. 

Our model is flexible and can be applied to other particle–surface combinations. For example, it can describe energetic neutral hydrogen atoms emitted from the lunar surface by replacing $\mathrm{P}^-$ with $\mathrm{P}^0$. Data from instruments such as ASAN on Chang’e-4 provide immediate opportunities to apply this model in other contexts.

\FloatBarrier

\begin{acknowledgements}
The Negative Ions on the Lunar Surface (NILS) instrument was developed by the Swedish Institute of Space Physics (IRF) in Kiruna, Sweden, on behalf of the European Space Agency (ESA). It was supported by the ESA grant No. 3-17483/22/NL/DB. Activities at NSSC were supported by the National Natural Science Foundation of China (NSFC), grant No. 42441807.
\\
NILS data is available from the European Space Agency's Planetary Science Archive (PSA) under doi:10.57780/esa-jw5mh1u.
\\
For solar wind parameters we acknowledge the use of ARTEMIS-P2 (THEMIS-C) data via NASA/GSFC's Space Physics Data Facility's OMNIWeb (or CDAWeb) service, and OMNI data.
\\
All posterior distributions were obtained through NUTS sampling \citep{Hoffman_2011} implemented in the PyMC Python package \citep{PYMC} with a NumPyro back-end \citep{phan2019composable, bingham2019pyro} for JAX-based computation. Some Equations were simplified through the use of PySR, a symbolic regression Python package \citep{Cranmer2023}.
\\
The authors used large language models to assist with code writing and improve the clarity of the English language.
\end{acknowledgements}

\bibliographystyle{bibtex/aa}
\bibliography{bibtex/biblio}

\begin{appendix}

\onecolumn
\section{Inference summary}

\begin{figure*}[ht!]
    \centering
     \resizebox{\textwidth}{\textwidth}
    {\includegraphics{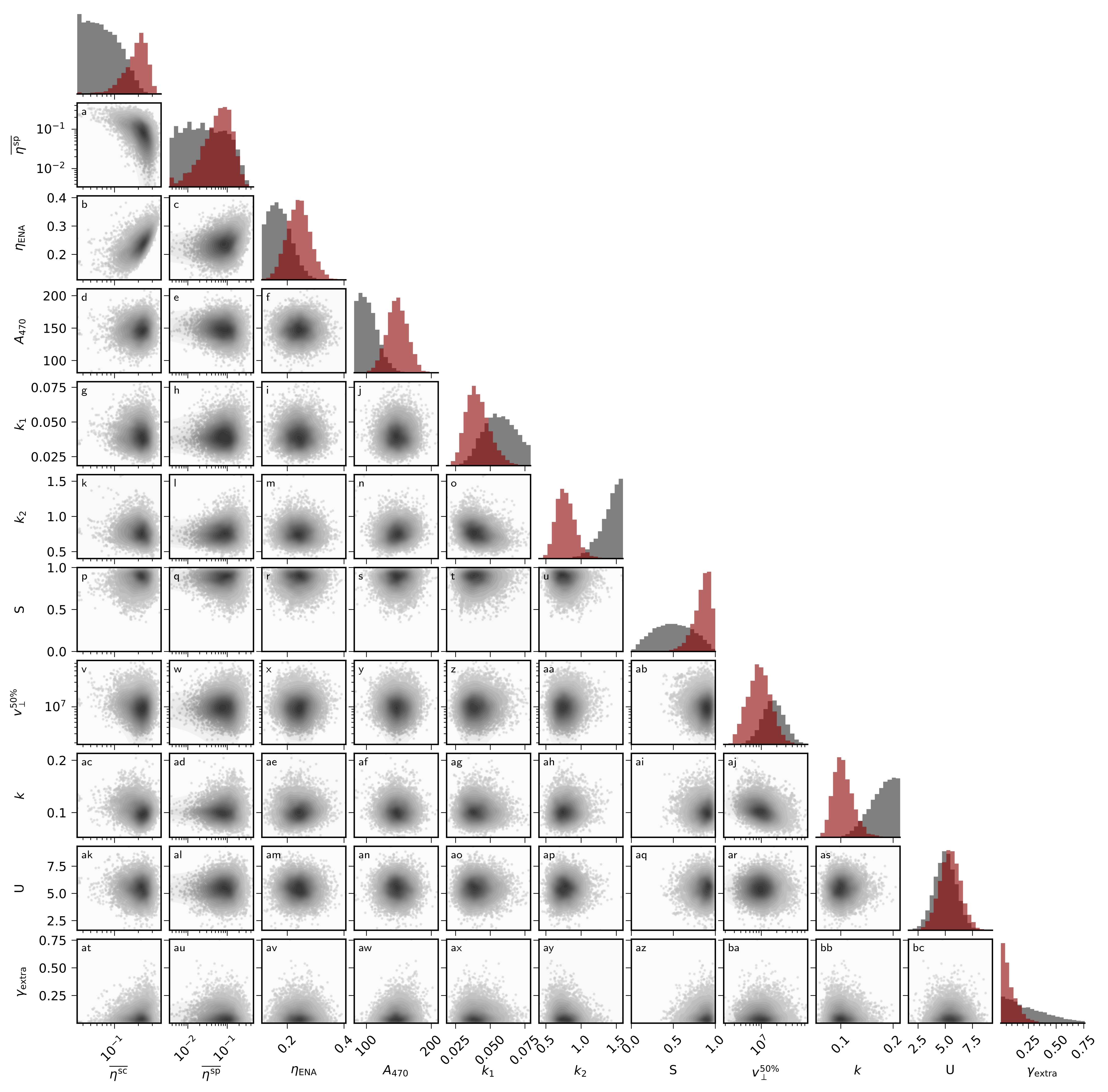}}
     \caption{Pair-plot of the joint posterior distribution $\mathbb{P}\left(\mathcal{J}_\mathrm{H}^-\mid\vec{\mathcal{C}}\right)$. Diagonal panels show the marginal posterior distributions (red) alongside the prior distributions (gray) for each model parameter. Note that prior distributions may be cropped to fit the panel x-axis range. Off-diagonal panels display posterior samples as points, with a faint overlay of a Gaussian kernel density estimate to illustrate the distribution more clearly. An alphabetical label is added to each panel for referencing.}
      \label{fig:pair_plot}
\end{figure*}

\FloatBarrier 
\twocolumn

\section{Regolith as a rough surface}
\label{appendix:regolith_rough_surface}

\subsection{Microscopic polar emission angle}

\citet{Szabo_2022a} present a statistical model for the distribution of inclination angles of a random Gaussian rough surface, characterized by a single roughness parameter, $w$. From this parameter, the mean surface inclination can be derived as \citep[Eq.~4]{Szabo_2022a}

\begin{equation}
\delta_m \;\left[\mathrm{rad}\right] \equiv \dfrac{\pi}{2} \exp\left[\frac{1}{2w^2}\right] \mathtt{erfc}\left[\frac{1}{w\sqrt{2}}\right]\;,
\label{eq:mean_inclination_angle}
\end{equation}

where $\mathtt{erfc}$ denotes the complementary error function. \citet{Broetzner2025} estimated the mean inclination angle of an Apollo 16 regolith sample to be $\delta_m = 27.7^\circ$. \citet{Helfenstein1999} report larger mean slopes of approximately $40^\circ$ at the 0.1\,mm scale, but with a significant uncertainty of about $20^\circ$ ($1\sigma$). For this reason, we use the first estimate $\delta_m = 27.7^\circ$ to describe the roughness of the lunar surface observed by NILS, and solve Eq.~\ref{eq:mean_inclination_angle} for $w$, obtaining $w\approx 0.45$. 

\begin{figure}[ht!]
    \centering
    \includegraphics[width=1\linewidth]{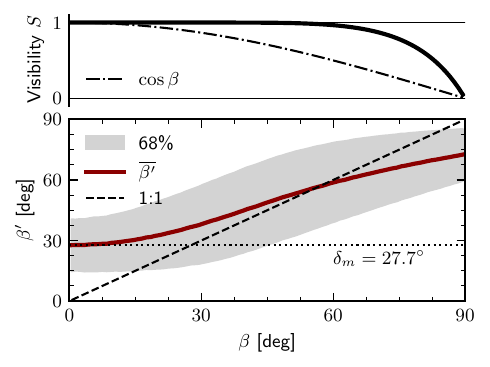}
    \caption{Mapping between the macroscopic polar emission angle $\beta$ and the microscopic polar emission angle $\beta'$ for a rough surface of mean inclination angle of $\delta_m=27.7^\circ$ (dotted line). The thick red line shows the average $\beta'$ for any $\beta$, along with a 68\% percentile range (gray shaded region), the unity line (dashed line).}
    \label{fig:polar_emission_angle}
\end{figure}

We construct a mapping between the macroscopic polar emission angle $\beta$ and the microscopic angle $\beta'$ using \citet[Eq.~C4]{Szabo_2022a}:

\begin{equation}
\cos\beta = \frac{\cos\beta' - \sin\beta'\cos\phi'\sqrt{p^2 + q^2}}{\sqrt{1 + p^2 + q^2}}\;,
\end{equation}

where $(p,q)$ are the surface slopes \citep[see Fig.~A1]{Szabo_2022a}, and $\phi'$ is the microscopic azimuthal emission angle. To obtain $\beta'$ from this relation, we sample $(p,q)$ from the joint normal distribution $(p,q) \sim \mathcal{N}(0, w^2)$ \citep[Eq.~A2]{Szabo_2022a}, and uniformly sample the macroscopic and microscopic angles: $\phi' \sim \mathtt{Uniform}(0, 2\pi)$ and $\cos\beta \sim \mathtt{Uniform}(0, 1)$. The resulting mapping is shown in Fig.~\ref{fig:polar_emission_angle}, with Eq.~\ref{eq:average_polar_emission_angle_mapping} drawn as a thick red line.

\subsection{Shadowing}

In the case of rough surfaces, a sputtered particle that leaves the surface may be redeposited on its way out. We describe the probability that a particle, leaving the surface at a macroscopic emission polar angle $\beta$, impact the surface again and be therefore unobserved by the instrument as the "visibility" probability $S$, defined as \citep[Eq.~9, 13, C5]{Szabo_2022a}:

\begin{subequations}
    \begin{align}
        S &= \dfrac{1}{1+\Lambda}\text{ , with }\\
        \Lambda &= \dfrac{w}{\cot\beta\sqrt{2\pi}} \exp\left(\dfrac{-\cot^2\beta}{2w^2}\right)-\dfrac{1}{2}\mathtt{erfc}\left(\dfrac{\cot\beta}{w\sqrt{2}}\right)\;.
    \end{align}
    \label{eq:visibility}
\end{subequations}

The visibility probability is given for $w=0.45$ in Fig.~\ref{fig:polar_emission_angle}.

\section{Instrument model}
\label{appendix:instrument_model}

The instrument reports a number of counts $\mathcal{C}$ per measurement time $\tau$. This section describes the conversion between the count rate $\mathcal{C}/\tau$ and the differential number flux $\mathcal{J}$.

The instrument scans the phase-space incrementally. We label each observation of the phase-space by a velocity vector $\vec{\mathrm{u}}$, as defined in \citet[][Eq.~5]{CanuBlot2025}. From \citet[][Eq.~55 and 58]{CanuBlot2025}, we write

\begin{equation}
A\left(\vec{\mathrm{u}}\right)\kappa\left(\vec{\mathrm{u}}\right) \approx \dfrac{2\xi_u}{\cos\omega_u} \mathcal{GF}_0\left(\vec{\mathrm{u}}\right)\eta\left(\vec{\mathrm{u}}\right)\;,
\label{eq:A_kappa}
\end{equation}

where $A \; [\mathrm{cm}^2]$ is the effective area of the instrument, $\mathcal{GF}_0 \; [\mathrm{cm}\;\mathrm{sr\;\mathrm{eV/eV}}]$ is the absolute geometric factor of the instrument \citep[][Eq.~73]{CanuBlot2025}, $\kappa \; [\mathrm{sr}\;\mathrm{eV/q}]$ is a scaling factor \citep[Eq.~10]{CanuBlot2025}, $\eta$ is the detection probability \citep[][Sect.~4.7.2]{CanuBlot2025}, and $(\xi_u, \omega_u)$ are the average energy-per-charge $[\mathrm{eV/q}]$ and elevation of a measurement $\vec{\mathrm{u}}$, respectively. The elevation angle is historically noted $\theta$ but to avoid conflicting notation, we name it $\omega$. The approximation in Eq.~\ref{eq:A_kappa} comes from the following simplification (with an average error of 1\% over the instrument coverage)

\begin{equation}
    \cos\omega_u \approx \iint \mathcal{R}_{\Omega}\left(\hat{\vec{\mathrm{u}}}, \omega, \alpha\right) \cos^2\alpha\cos\omega \mathrm{d}\alpha \mathrm{d}\omega\;,
\end{equation}

where $\mathcal{R}_\Omega \; [\mathrm{1/sr}]$ is the angular response of the instrument and $\alpha$ the azimuthal angle. 

We replace $A\left(\vec{\mathrm{u}}\right)\kappa\left(\vec{\mathrm{u}}\right)$ in \citet[Eq. 57]{CanuBlot2025} by the approximation of Eq.~\ref{eq:A_kappa}, make the assumption that the differential flux does not posses any structure smaller than the energy response of the instrument, and note that $\mathcal{J} = 2Ef/m$, with $f$ the phase-space density of a species of mass $m$, and we obtain

\begin{subequations}
    \begin{align}
        \lambda\left(\vec{\mathrm{u}}\right) &= \mathcal{K}\left(\vec{\mathrm{u}}\right) \iint \mathcal{J}\left(\xi_u,\omega,\alpha\right) \mathcal{R}_{\Omega}\left(\hat{\vec{\mathrm{u}}},\omega,\alpha\right)\cos^2\alpha\cos\omega \mathrm{d}\alpha \mathrm{d}\omega\;,\\
        \mathcal{K}\left(\vec{\mathrm{u}}\right)&\equiv 2\tau\dfrac{\xi_u}{\cos\omega_u}\eta\left(\mathrm{u}\right)\mathcal{GF}_0\left(\vec{\hat{\mathrm{u}}}\right).
    \end{align}
    \label{eq:counts_model_1}
\end{subequations}

We now make the assumption that the differential flux varies little over the azimuthal coverage of the angular response of the instrument. This gives

\begin{subequations}
    \begin{align}
        &\lambda\left(\vec{\mathrm{u}}\right) = \mathcal{K}\left(\vec{\mathrm{u}}\right) \int_{3\pi/2}^{2\pi} \mathcal{J}\left(\xi_u,\omega,\alpha_u\right) \cdot \mathcal{\tilde{R}}_{\Omega}\left(\mathbf{\hat{\vec{\mathrm{u}}}},\omega\right)\cos\omega \mathrm{d}\omega\,,\\
        &\mathcal{\tilde{R}}_{\Omega}\left(\hat{\vec{\mathrm{u}}},\omega\right)\equiv\int\mathcal{R}_{\Omega}\left(\hat{\vec{\mathrm{u}}},\omega,\alpha\right)\cos^2\alpha \mathrm{d}\alpha.
    \end{align}
    \label{eq:counts_model_2}
\end{subequations}

The integral in Eq. \ref{eq:counts_model_2} is defined in the instrument frame (see \citet[Sect. 4.1.3]{CanuBlot2025} for more details), and we now express it using the emission angles $\left(\beta, \phi\right)$ (see Fig.~\ref{fig:all_angles}) using the transformation defined in \citet[Eqs. 14, 15, and 16]{Wieser2025}.

\begin{equation}
    \boxed{\lambda\left(\vec{\mathrm{u}}\right) = \mathcal{K}\left(\vec{\mathrm{u}}\right) \int_0^{\pi/2} \mathcal{J}\left(\xi_u,\beta,\phi_u\right) \cdot \mathcal{\tilde{R}}_{\Omega}\left(\hat{\vec{\mathrm{u}}},\beta-\dfrac{\pi}{2}\right)\sin \beta \mathrm{d}\beta}\;,
    \label{eq:instrument_model}
\end{equation}

with $\phi_u = f_2\left(\omega_u, \alpha_u;\mathrm{SA}\right)$, where $f_2$ is defined in \citet[Eq. 13]{Wieser2025} and $\alpha_u=-0.752^\circ$ \citep[Eq. 16]{CanuBlot2025}. We neglect the dependency over $\omega$, and approximate $\phi_u \approx \overline{\phi} = f_2\left(\alpha_u; \overline{\mathrm{SA}}\right) \approx 213^\circ$, where $\overline{\mathrm{SA}} \approx 34^\circ$ is the average Solar Azimuth angle over the mission. 

\section{Bayesian inference}
\label{appendix:bayesian}

The NILS dataset consists of 302 minutes of observations, with the instrument viewing the lunar surface for about half of that time. The near-surface environment is dominated by secondary electrons, photoelectrons, and UV radiation, all of which complicate the study of the weaker negative ion signal. We use Bayesian inference to extract and study the signal.

We first define the set of observed counts as $\vec{\mathcal{C}}$, that is, the number of counts observed in the time-energy-angular-mass matrix. We assume that the observed counts are drawn from a Poisson (\texttt{Pois}) distribution such that

\begin{equation}
    \mathbb{P}\left(\vec{\mathcal{C}} \mid \vec{\Lambda}, \vec{\mathrm{F}}\right) = \mathtt{Pois}\left(\vec{\Lambda} \cdot \vec{\mathrm{F}}\right)\;,
    \label{eq:counts_model}
\end{equation}

where $\vec{\Lambda} \equiv \left[\lambda_{\mathrm{H}^-}, \vec{\lambda_\mathrm{nuis}}\right]$, with $\lambda_{\mathrm{H}^-}$ the rate of negative hydrogen ions and $\vec{\lambda_\mathrm{nuis}}$ the contribution of all other nuisance signals--namely, electrons, oxygen ions and UV. The matrix $\vec{\mathrm{F}}$ defines the mass response of the instrument, as given by \citet[Sect.~4.6]{CanuBlot2025}. The matrix $\vec{\Lambda}$ thus defines the signal rates over the time-energy-angular space.

We wish to infer knowledge about the negative hydrogen ion differential flux $\mathcal{J}_\mathrm{H}^-$. We link the rate $\lambda_{\mathrm{H}^-}$ with $\mathcal{J}_\mathrm{H}^-$ using the instrument response defined by Eq.~\ref{eq:instrument_model}. We thereafter simplify the notation by introducing $\lambda\equiv\lambda_{\mathrm{H}^-}$ and $\mathcal{J} \equiv \mathcal{J}_{\mathrm{H}}^-$, and obtain the deterministic relation

\begin{equation}
    \mathbb{P}\left(\lambda \mid \mathcal{J}\right) = \delta\left[\lambda - \lambda_{\mathrm{model}}\left(\mathcal{J}\right)\right]\;,
    \label{eq:flux_model}
\end{equation}

where the relation $\lambda_\mathrm{model}\left(\mathcal{J}\right)$ is defined by Eq.~\ref{eq:instrument_model}. We introduce the statistical distribution of the differential flux conditional on the observed data (thereafter called \emph{posterior distribution}), $\mathbb{P}\left(\mathcal{J}\mid\vec{\mathcal{C}}\right)$. We apply Bayes’ theorem and express the probability distribution in proportional form by omitting the marginal likelihood

\begin{equation}
    \mathbb{P}\left(\mathcal{J}\mid\vec{\mathcal{C}}\right) \propto \mathbb{P}\left(\mathcal{J}\right) \mathbb{P}\left(\vec{\mathcal{C}} \mid \mathcal{J}\right)\;. 
    \label{eq:posterior_dist}
\end{equation}

We introduce the rate matrix $\vec{\Lambda}$ and the mass response of the instrument $\vec{\mathrm{F}}$ into $\mathbb{P}\left(\vec{\mathcal{C}} \mid \mathcal{J}\right)$, giving:

\begin{equation}
    \mathbb{P}\left(\vec{\mathcal{C}} \mid \mathcal{J}\right) = \iint \mathbb{P}\left(\vec{\mathcal{C}}, \lambda, \vec{\lambda_\mathrm{nuis}}, \vec{\mathrm{F}} \mid \mathcal{J}\right) \mathrm{d}\lambda \mathrm{d}\vec{\lambda_\mathrm{nuis}} \mathrm{d}\vec{\mathrm{F}}\;.
\end{equation}

From the assumptions that (i) $\lambda$ and $\vec{\lambda_\mathrm{nuis}}$ are mutually-independent; (ii) $\vec{\mathrm{F}}$ and $\vec{\lambda_\mathrm{nuis}}$ do not depend on $\mathcal{J}$; (iii) the uncertainty in the mass response of the instrument can be neglected, we obtain

\begin{equation}
    \mathbb{P}\left(\vec{\mathcal{C}} \mid \mathcal{J}\right) = \int \mathbb{P}\left(\vec{\mathcal{C}} \mid \lambda, \vec{\lambda_\mathrm{nuis}}, \vec{\mathrm{F}}\right) \mathbb{P}\left(\lambda \mid \mathcal{J}\right) \mathbb{P}\left(\vec{\lambda_\mathrm{nuis}} \right) \mathrm{d}\lambda \mathrm{d}\vec{\lambda_\mathrm{nuis}}\;.
\end{equation}

From Eqs.~\ref{eq:counts_model} and \ref{eq:flux_model}, and integrating over $\lambda$, we obtain

\begin{equation}
    \mathbb{P}\left(\mathcal{J}\mid\vec{\mathcal{C}}\right) \propto \mathbb{P}\left(\mathcal{J}\right) \int \mathtt{Pois}\left(\left[\lambda_\mathrm{model}\left(\mathcal{J}\right), \vec{\lambda_\mathrm{nuis}}\right] \cdot \vec{\mathrm{F}}\right) \mathbb{P}\left(\vec{\lambda_\mathrm{nuis}}\right) \mathrm{d}\vec{\lambda_\mathrm{nuis}}\;. 
\end{equation}

We used a Jeffreys prior (non-informative) on $\mathbb{P}\left(\vec{\lambda_\mathrm{nuis}}\right)$. The prior $\mathbb{P}\left(\mathcal{J}\mid\vec{\mathcal{C}}\right)$ is described through the priors in Sect.~\ref{sec:prior_knowledge}, the Table~\ref{tab:priors}, and the model in Eq.~\ref{eq:full_flux_model_applied_to_NILS}.

\section{Normalization of scattering energy model}
\label{appendix:scattered_norm_cst}

The model for the energy distribution of scattered particles is computationally expensive, especially Eq.~\ref{eq:scattered_model_straggling}, which evaluates the function $\mathcal{J}_\mathrm{E}^\mathrm{sc}$ $N$-times to marginalize it over the inelastic energy loss. We give below an alternative definition of the normalization factor $\mathtt{n}$, obtained from the non-dimensionalization of the distribution $\mathcal{J}_\mathrm{E}^\mathrm{sc}$.

\subsection{Non-dimensionalization}

We reduce the dimensionality of the problem by normalizing over the incident energy $E_\mathrm{in}$. We introduce the dimensionless variables 

\begin{subequations}
    \begin{align}
        e &= E/E_\mathrm{in}\;,\\
        u &= U/E_\mathrm{in}\;,\\
        \Delta_\epsilon' &= \Delta_\epsilon/E_\mathrm{in}\;,\\
        \text{giving }\zeta &= \dfrac{A}{e+u}\;,\\
        \text{with }A &= p^2 \left(1+u\right)-\Delta_\epsilon'\;.
    \end{align}
\end{subequations}

We obtain the energy distribution in reduced form

\begin{equation}
    E_\mathrm{in} \cdot \mathcal{J}_E^{\mathrm{sc}} = \dfrac{2\kappa e}{\left(e+u\right)^2} f\left(x\right)\;,
\end{equation}

with $x$ and $f$ defined as in Eq.~\ref{eq:scattered_model}. Noting that 

\begin{subequations}
    \begin{align}
        \mathrm{d}E &= \dfrac{E_\mathrm{in}A}{\kappa \exp\left(x/\kappa\right)}\mathrm{d}x\;,\\
        e&=\dfrac{A}{\exp\left(x/\kappa\right)}-u\;,
    \end{align}
\end{subequations}

we can write the normalization factor $\mathtt{n} = \int_0^\infty \mathcal{J}_E^{\mathrm{sc}} \mathrm{d}E$, independent on $E_\mathrm{in}$, as:

\begin{equation}
    \mathtt{n}\left(u,\Delta_\epsilon'\right) = 2 \int_0^{\,\kappa\ln \left(A/u\right)} \left[1-\dfrac{u}{A}\exp\left(x/\kappa\right)\right]f\left(x\right)\mathrm{d}x\;.
\end{equation}

\subsection{Analytical approximation}

For the special case of $\left(p=0.97, \kappa=7.3\right)$, corresponding to a proton impacting lunar regolith at 300 km/s, that is $M_\mathrm{P}=1\,\mathrm{amu}, M_\mathrm{S}=21.9\,\mathrm{amu}$, and $m=0.57$, we approximate the function $\mathtt{n}\left(u,\Delta_\epsilon'\right)$ by

\begin{equation}
    \mathtt{n}\left(u,\Delta_\epsilon'\right) \approx \left(\dfrac{\ln\Delta_\epsilon' - 2.193 \times 10^{-1}}{\ln u\cdot\ln\Delta_\epsilon'}\right) - \dfrac{1.262 \times 10^{-2}}{\ln\Delta_\epsilon'} +9.010 \times 10^{-1}\;.
\end{equation}

The approximation is valid for $\Delta_\epsilon' \in\left[1\times10^{-3}, 0.95\right]$ and $u\in\left[1\times10^{-3}, 2\times10^{-2}\right]$, which covers most of the possible surface binding energies, inelastic losses, and solar wind energies. Within this space, the approximation has a mean relative error of 0.4\%, and loses accuracy (relative error of about 10\%) for large $\Delta_\epsilon' > 0.9$.

\section{Extra content}

\subsection{Definition of priors}

We define here all prior distributions not mentioned in Sect.~\ref{sec:prior_knowledge}.

\begin{table}[h!]
\caption{Definition of prior knowledge}
\label{tab:priors}
\centering
\renewcommand{\arraystretch}{1.3}
\setlength{\tabcolsep}{10pt}
\begin{tabular}{@{}lll@{}}
\hline\hline
Parameter & Distribution & Constraint \\ \midrule

$\mathrm{S}$ & $\mathtt{Beta}\left(\alpha=2, \beta=2\right)$ &  \\ 
$v_\perp^{50\%}$ & $\mathtt{LogNormal}\left(\mu=16.9, \sigma=0.5\right)$ & \\ 
$k$ & $\mathtt{LogNormal}\left(\mu=-1.51, \sigma=0.25\right)$ & \\ 
\hline

$\mathrm{A}_{470}$ & $\mathtt{Normal}\left(\mu=94.6, \sigma=20.0\right)$ & $\mathrm{A}_{470} > 0$ \\ 
$k_1$ & $\mathtt{LogNormal}\left(\mu=-2.83, \sigma=0.25\right)$ & \\ 
$k_2$ & $\mathtt{LogNormal}\left(\mu=0.71, \sigma=0.25\right)$ & \\ 
\hline

$\gamma_\mathrm{extra}$ & $\mathtt{Beta}\left(\alpha=1, \beta=3\right)$ & \\ 
$U$ & $\mathtt{Normal}\left(\mu=5, \sigma=1\right)$ & $U\ge1$ \\ 
$r$ & $\mathtt{Uniform}\left(0, 1\right)$ & \\ 
\hline

$\epsilon_i^\mathrm{sc}$ & $\mathtt{Normal}\left(\mu=0, \sigma=1\right)$ & \\ 
$\epsilon_i^\mathrm{sp}$ & $\mathtt{Normal}\left(\mu=0, \sigma=1\right)$ & \\ 
$\sigma_\mathrm{sc}$ & $\mathtt{HalfNormal}\left(\sigma=1\right)$ & \\ 
$\sigma_\mathrm{sp}$ & $\mathtt{HalfNormal}\left(\sigma=1\right)$ & \\ 
\hline
$\vec{\lambda_\mathrm{nuis}}$ & $\propto 1/\sqrt{\vec{\lambda_\mathrm{nuis}}}$ & $\vec{\lambda_\mathrm{nuis}} > 0$\\
\bottomrule
\end{tabular}
\end{table}

The dependence of the scattering angular emission distribution $\mathcal{J}_{\Omega}^\mathrm{sc}\left(\beta_i,\overline{\phi}\right)$ and the sputtering angular emission distribution $\mathcal{J}_{\Omega}^\mathrm{sp}\left(\beta_i,\overline{\phi}\right)$ over the macroscopic emission polar angle $\beta_i$ is discretized in 11 linearly-spaced angular intervals, indexed $i=1, \ldots, 11$. We used a hierarchical, non-centred log-normal parametrization defined as (for the sputtered process, for example)

\begin{equation}
    \mathcal{J}_{\Omega}^\mathrm{sp}\left(\beta_i,\overline{\phi}\right) = \exp\left[\ln\left[\mathcal{J}_{\Omega}^\mathrm{sp, \, prior}\left(\beta_i,\overline{\phi}\right)\right] + \sigma^\mathrm{sp} \times \epsilon_i^\mathrm{sp}\right]\;.
\end{equation}

The choice of a hierarchical parametrization allows for smooth and weakly informative variation across angular bins, while a non-centred parametrization reduces the correlation between the per-bin angular yield and the scale parameter $\sigma_\mathrm{sp}$ and $\sigma_\mathrm{sc}$\citep{Papaspiliopoulos2007}.

We systematically verified the choice of priors through predictive prior modelling. We studied heuristically the sensitivity of the inference on the choice of priors. In all cases, the main conclusions of this study remained robust to reasonable variations in the prior assumptions.

\subsection{Posterior summary of angular yields}
Table \ref{tab:weights} summarizes the marginal posterior distributions of the scattering and sputtering angular yields.

\begin{table*}[h!]
\caption{Summary of marginal posterior distributions for the sputtering and scattering angular yields.}
\label{tab:weights}
\centering
\renewcommand{\arraystretch}{1.1}
\setlength{\tabcolsep}{10pt}
\begin{tabular}{l|llll|llll}
\hline\hline
\tablefootmark{a}$\beta_i$ [deg] & \multicolumn{4}{c|}{$\overline{\eta^\mathrm{sp}}\mathcal{J}_\Omega^\mathrm{sp}\left(\beta_i, \overline{\phi}\right) \; [1/\mathrm{sr}]$} & \multicolumn{4}{c}{$\overline{\eta^\mathrm{sc}}\mathcal{J}_\Omega^\mathrm{sc}\left(\beta_i, \overline{\phi}\right) \; [1/\mathrm{sr}]$} \\
 & \texttt{MAP} & median & \multicolumn{2}{c|}{68\% \texttt{HDI}} & \texttt{MAP} & median & \multicolumn{2}{c}{68\% \texttt{HDI}} \\
 & $\times 10^{2}$ & $\times 10^{2}$ & $\mathrm{low}\times 10^{3}$ & $\mathrm{high}\times 10^{2}$ & $\times 10^{2}$ & $\times 10^{2}$ & $\mathrm{low}\times 10^{3}$ & $\mathrm{high}\times 10^{2}$\\
\hline
39 & 3.24 & 2.73 & 4.5 & 19.3 & 2.3 & 2.26 & 12.4 & 3.78 \\
44 & 3.25 & 3.9 & 3.43 & 39 & 2.26 & 2.46 & 13.6 & 4.56 \\
49 & 52.4 & 7.45 & 12.9 & 89.6 & 2.14 & 2.09 & 12.4 & 3.33 \\
54 & 34.5 & 6.56 & 15.6 & 80 & 1.93 & 1.8 & 11.8 & 3.08 \\
59 & 32.8 & 23 & 87.5 & 73.9 & 1.85 & 1.9 & 12.5 & 3.07 \\
63 & 2.38 & 1.86 & 4.75 & 13.4 & 1.69 & 1.72 & 11.4 & 2.73 \\
68 & 2.53 & 1.29 & 3.56 & 8.45 & 1.53 & 1.68 & 10.4 & 2.65 \\
73 & 1.1 & 0.802 & 2.28 & 4.8 & 1.29 & 1.43 & 9.32 & 2.32 \\
78 & 0.642 & 0.405 & 0.959 & 1.81 & 1.01 & 1.03 & 6.54 & 1.53 \\
83 & 0.401 & 0.361 & 0.784 & 2.13 & 0.686 & 0.682 & 4.18 & 1.07 \\
88 & 0.144 & 0.163 & 0.219 & 1.18 & 0.351 & 0.333 & 1.95 & 0.558 \\
\hline
\end{tabular}
\tablefoot{As a reminder, the angular yields are given for $\overline{\phi} = 213^\circ$, a SZA ranging from $47^\circ$ to $52^\circ$, and a solar wind proton energy ranging from $440$\,eV to $502$\, eV. The \texttt{MAP} and \texttt{HDI} are calculated in log-space and exponentiated back to linear-space.\\
\tablefoottext{a}{$\beta_i$ is the centre of the angular intervals, with a constant width of 5 deg.}
}
\end{table*}

\end{appendix}
\end{document}